\pdfoutput=1
\documentclass[a4paper,11pt]{article}
\usepackage{jheppub}

\usepackage{amsmath,amssymb,amsfonts}
\usepackage{tikz}
\usepackage[all]{xy}
\usetikzlibrary{shapes,arrows}
\tikzstyle{block} = [rectangle, draw, text width=7em, text centered, rounded corners, minimum height=3em]
\usepackage{graphicx,color}
\usepackage{tikz}
\usepackage{tikz-cd}
\usetikzlibrary{cd}

\usepackage{caption}
\usepackage{subcaption}

\usepackage{tikz}
\usetikzlibrary{shapes}
\usetikzlibrary{positioning}
\usetikzlibrary{chains}
\usetikzlibrary{arrows,fit,decorations.pathreplacing, shapes.misc}
\usetikzlibrary{decorations.markings}
\tikzstyle{every picture}+=[remember picture]
\tikzstyle{na} = [baseline=-.5ex]
\usepackage{todonotes}
\usepackage{tikz-3dplot}

\def\bZ{\mathbb{Z}}
\def\bR{\mathbb{R}}

\def\bC{\mathbb{C}}
\def\bP{\mathbb{P}}

\def\cN{\mathcal{N}}
\def\cO{\mathcal{O}}

\newcommand{\Tr}{{\rm Tr}}
\def\tr{\mathop{\mathrm{tr}}}

\title{\boldmath MSW-type compactifications of 6d $(1,0)$ SCFTs on 4-manifolds}
\author[\ddagger]{Jin Chen,}
\author[\ast]{Zhuo Chen,}
\author[\ast,\dag]{Wei Cui,}
\author[\dag,\ast]{Babak Haghighat}
\affiliation[\ast]{Yanqi Lake Beijing Institute of Mathematical Sciences and Applications (BIMSA), Huairou District, Beijing 101408, P. R. China}
\affiliation[\dag]{Yau Mathematical Sciences Center, Tsinghua University, Beijing, 100084, China}
\affiliation[\ddagger]{Department of Physics, Xiamen University, Xiamen, 361005, China}

 \emailAdd{zenofox@gmail.com}
 \emailAdd{zhuo2012@vt.edu}
 \emailAdd{cwei@bimsa.cn}
 \emailAdd{babakhaghighat@tsinghua.edu.cn}

\abstract{In this work, we study compactifications of 6d $(1,0)$ SCFTs, in particular those of conformal matter type, on K\"ahler 4-manifolds. We show how this can be realized via wrapping M5 branes on 4-cycles of non-compact Calabi-Yau fourfolds with ADE singularity in the fiber. Such compactifications lead to domain walls in 3d $\mathcal{N}=2$ theories which flow to 2d $\mathcal{N}=(0,2)$ SCFTs. We compute the central charges of such 2d CFTs via 6d anomaly polynomials by employing a particular topological twist along the 4-manifold. Moreover, we study compactifications on non-compact 4-manifolds leading to coupled 3d-2d systems. We show how these can be glued together consistently to reproduce the central charge and anomaly polynomial obtained in the compact case. Lastly, we study concrete CFT proposals for some special cases.}

\begin{document} 
\maketitle
\flushbottom

\section{Introduction}

The existence of six-dimensional superconformal field theories (SCFTs) has initiated a classification program for constructions of lower $d$-dimensional quantum field theories in terms of geometries of $(6-d)$-dimensional manifolds on which the 6d theories are compactified \cite{Gaiotto:2009we,Gaiotto:2009hg,Argyres:2007cn,Argyres:2007tq,Alday:2009aq,Benini:2009mz,Dimofte:2010tz,Dimofte:2011jd,Dimofte:2011ju,Gadde:2013sca,Gukov:2016gkn,Zafrir:2015rga,Jefferson:2018irk,Bhardwaj:2019fzv,Braun:2021lzt,Gaiotto:2015una,Razamat:2016dpl,Bah:2017gph,Kim:2017toz,Kim:2018bpg,Kim:2018lfo,Razamat:2018gro,Chen:2019njf,Razamat:2019mdt,Cecotti:2011iy,Cordova:2013cea,Yagi:2013fda,Lee:2013ida,Gukov:2015sna,Cordova:2016cmu,Gukov:2017kmk,Cho:2020ljj,Apruzzi:2016nfr,Gukov:2018iiq, Pasquetti:2019hxf,Assel:2022row}. Within this setup, compactifications of 6d (2,0) SCFTs, realized by $N$ parallel M5-branes, along various 4-manifolds have been a very fruitful approach to construct two-dimensional CFTs with chiral supersymmetries \cite{Maldacena_1997,Gaiotto:2006wm,Gadde:2013sca,Dedushenko:2017tdw,Feigin:2018bkf}. The amount of supersymmetry of the resulting two-dimensional theories depends on the choice of different topological twists of the underlying 6d SCFT along the 4-manifolds. This way, different supersymmetry algebras, namely $\mathcal{N}=(0,2)$ or $\mathcal{N}=(0,4)$, can be realized when the M5 branes are wrapping a 4-cycle inside a $G_2$ manifold or a Calabi-Yau threefold, respectively \cite{Gukov:2018iiq}. 

A direct but intriguing generalization of the above is to consider compactifications of 6d $(1,0)$ SCFTs. Due to a much richer classification of 6d $(1,0)$ theories \cite{Witten:1995ex, Witten:1995zh, Witten:1995gx, Seiberg:1996vs, Morrison:1996pp, Seiberg:1996qx, Hanany:1997gh, Morrison:2012np, Heckman:2013pva, DelZotto:2014hpa, Heckman:2014qba, Heckman:2015bfa}, it is expected that their compactification on various manifolds will lead to a far vaster landscape of lower dimensional quantum field theories. Recently, the investigation of compactifications of such theories on 4-manifolds has been initiated \cite{Apruzzi:2016nfr,Gukov:2018iiq}. Again, there are two different choices for the topological twist upon compactification which result in 2d $\mathcal{N}=(0,1)$ or $(0,2)$ theories, respectively. The latter choice is only possible for K\"ahler 4-manifolds.

In this work we will continue the investigation of the compactifications of 6d $(1,0)$ SCFTs on 4-manifolds where we will be mainly focusing on the class of conformal matter theories \cite{DelZotto:2014hpa} and the twist which leads to $\mathcal{N}=(0,2)$ supersymmetry in two dimensions. We will argue that this twist is naturally realized in fivebrane worldvolumes which wrap 4-cycles in Calabi-Yau fourfolds with ADE singularities in their fiber. This is analogous to the approach of Maldacena, Strominger and Witten (MSW) \cite{Maldacena_1997} who wrapped the $(2,0)$ theory of M5 branes on 4-cycles of Calabi-yau threefolds. Fivebranes wrapped on K\"ahler 4-cycles of Calabi-Yau fourfolds give rise to domain walls in three dimensions \cite{Gukov:1999ya} and the goal is to study their physics. The reduction of the M-theory effective action along the non-compact Calabi-Yau fourfold with appropriate $G$-flux turned on then leads to a 3d $\mathcal{N}=2$ theory which in the case of A-type singularities specializes to $SU(k)$ Chern-Simons theory. One can then proceed to count vacua on both sides of the 2d domain wall to obtain the degrees of freedom at the interface. Moreover, we will compute anomaly polynomials and central charges of the resulting 2d theories by alternatively reducing the corresponding 6d anomaly polynomials along 4-manifolds. We then proceed to decompose the 4-manifold into non-compact 4-manifolds which are glued together to obtain a compact space along the lines discussed in \cite{Bonelli:2012ny,Feigin:2018bkf} for the 6d $(2,0)$ theory. We find that central charge expressions for 6d $(1,0)$ theories compactified on the non-compact patches can be obtained by a regularization procedure and add correctly together to reproduce the central charges of the compact 4-manifold. Moreover, we interpret the compactification on the non-compact space as a coupled 3d-2d system where the 2d $\mathcal{N}=(0,2)$ theory is viewed as the boundary of a 3d supersymmetric TQFT such that the combined system is free of anomalies. When two non-compact manifolds are glued together, their 2d boundary theories fuse to a new 2d theory and compactification on the compact 4-manifold can be viewed as a 3d theory on a slab. Finally, we study a series of specific compactifications of 6d $(2,0)$ theories by setting the regularization parameters we investigated before to certain discrete rational values. We find that the resulting central charges from anomaly polynomials precisely match with those of $W_N(m,n)$ minimal models and thus interpret them as boundary CFTs of 3d TQFTs with anyons corresponding to the primary fields of the 2d boundary theories. In addition, we consider some generic features of the compactifications of $N$ M5-branes probing $\mathbb{C}^2/\mathbb{Z}_k$ singularities, namely the $\mathcal N=(1,0)$ class $S_k$ theories, on K\"ahler manifolds, as a generalization of the $(2,0)$ setups. We study the scaling behaviour of the central charges of the mysterious 2d theories when taking both $N$ and $k$ to be large. We find that the scaling matches with the one of the $k$th para-Toda CFT of type $SU(Nk)$. Although the matching is only asymptotic, the correct scaling behaviour may be a hint that a particular modification of the $k$th para-Toda theory turns out to be the correct 2d CFT description for such compactifications. 

The organization of this paper is as follows. In Section \ref{sec:CY4}, we describe the Calabi-Yau fourfold backgrounds in M-theory which are relevant for this work and deduce the corresponding 3d theories for fourfolds with A-type singularities. This allows to perform a counting on the degrees of freedom of the domain walls in such theories. In Section \ref{sec:MSWtwist}, we describe the MSW twist of 6d $(2,0)$ and $(1,0)$ theories when compactified on K\"ahler 4-manifolds. Using anomaly polynomials of the 6d theories and geometric data, we compute the central charges of the corresponding 2d theories. In Section \ref{sec:gluing}, we study the compactifications along non-compact 4-manifolds and the resulting coupled 3d-2d systems. We further show how to reproduce the 2d central charges for a compact manifold obtained by gluing several such non-compact patches together. In Section \ref{sec:MTC}, we give concrete proposals for 2d CFTs obtained from compactifications. Finally, in Section \ref{sec:conclusions}, we present our conclusions.

\section{$CY_4$ background for M5 branes probing ADE singularities}
\label{sec:CY4}

M5 branes wrapping fourmanifolds can give rise to 2d theories in various different scenarios. Given an M-theory background where the fourmanifold is a co-associative cycle in a manifold with $G_2$ holonomy, the resulting 2d theory has $\mathcal{N}=(0,2)$ supersymmetry \cite{Gadde:2013sca}. This can be seen by noting that the $G_2$ background preserves $4$ supercharges in the orthogonal four spacetime dimensions. Since the M5 brane forms a half BPS string which is of co-dimension two there, the corresponding 2d worldvolume theory preserves $2$ supercharges. Now, there is exactly one topological twist on the fivebrane worldvolume which preserves these supercharges, namely the one which embeds an $SU(2)$ subgroup of the $SO(4)$ holonomy of the fourmanifold in question into an $SU(2)$ subgroup of the R-symmetry. In the case that the fourmanifold is K\"ahler, we have another choice for the topological twist. In that case the holonomy group is $U(2)$ and we can embed the $U(1)$ factor of it into a $U(1)$ subgroup of the $SO(5)$ R-symmetry, see Section \ref{sec:MSWtwist}. This twist will give rise to $\mathcal{N}=(0,4)$ supersymmetry on the remaining two orthogonal spacetime dimensions of the fivebrane. We will call this twist the MSW twist since it is naturally realized in an M-theory background where the M5 brane wraps a K\"ahler four-cycle $P$ inside a Calabi-Yau three-fold \cite{Maldacena_1997}. In order to decouple gravity, one takes the Calabi-Yau to be the anti-canonical bundle of $P$,
\begin{equation}
    CY_3 \equiv \mathcal{O}(-K_P) \longrightarrow P~,
\end{equation}
preserving $8$ supercharges in the remaining five orthogonal directions. The fivebrane forms a half-BPS string there and thus its worldvolume will preserve $4$ supercharges \cite{Alim:2010cf,Haghighat:2011xx,Haghighat:2012bm}. 

A similar two-fold choice is possible for 6d $\mathcal{N}=(1,0)$ SCFTs when wrapped on fourmanifolds. There will be one twist which preserves just one supercharge in the remaining two orthogonal directions, while in the case of K\"ahler manifolds, there will be another twist preserving two supercharges, see the discussion in Section \ref{sec:MSWtwist}. We again call it, in analogy to the $\mathcal{N}=(2,0)$ case of M5 branes, an MSW-type twist. For the purposes of this paper, the relevant $(1,0)$ theories will be M5 probing ADE singularities, known as conformal matter theories \cite{DelZotto:2014hpa}, and M5 branes on top of an M9 wall arising from M-theory on $S^1/\mathbb{Z}_2$ \cite{Horava:1995qa}. In both cases, the MSW-type twist can be naturally realized by embedding the four-cycle into a Calabi-Yau fourfold. The relevant fourfold can be constructed in two steps. As a first step, one mods out $\mathbb{C}^2$ by a discrete subgroup $\Gamma_G$ of $SU(2)$ where $G$ is of ADE type. The resulting space $\mathbb{C}^2/\Gamma_G$ has zero first Chern class and an ADE singularity at the origin. One then fibers this space over the K\"ahler manifold $P$ in such a way that the first Chern class of the normal bundle cancels the one of the tangent bundle of $P$. Practically, this can be achieved by an elliptic fibration with discriminant locus equal to $P$ \cite{Morrison:2014lca}. For example, in the case of $P= \mathbb{P}^2$, one first forms a compact base $\widetilde{P}_n$ that is a $\mathbb{P}^1$ bundle over $\mathbb{P}^2$ by projectivization of the line bundle $\mathcal{O}(-n H)$, where $H$ is the hyperplane class of $\mathbb{P}^2$. One then fibers an elliptic curve $\mathcal{E}$ over $\widetilde{P}_n$ in such a way that the total space is Calabi-Yau,
\begin{equation}
    CY_4 \quad \equiv \quad \begin{array}{ccc}\mathcal{E} & \hookrightarrow & X \\
                                            ~       &         ~          & \downarrow \\
                                            ~       &         ~          & \widetilde{P}_n
                      \end{array}.
\end{equation}
In order to obtain, for example, a $D_4$ singularity, one has to choose $n=6$ (see \cite{Morrison:2014lca} for details) and successively send the volumes of the elliptic fiber $\mathcal{E}$ and the $\mathbb{P}^1$ fiber of $\widetilde{P}_n$ to infinity. 

The so constructed Calabi-Yau background preserves four supercharges in M-theory and an M5 brane wrapping $P$ will break two of those, thus preserving two supercharges in the orthogonal two spacetime dimensions. On the worldvolume level, these are realized by the MSW-type twist giving rise to $\mathcal{N}=(0,2)$ supersymmetry in 2d. This 2d theory is realized as a domain-wall inside a 3d $\mathcal{N}=2$ theory. These domain-walls fractionate in the case of $D$- and $E$-type singularities \cite{DelZotto:2014hpa}.

\subsection{Counting Domain Walls}
\label{sec:DWcount}

In the following, we want to count the degrees of freedom associated to the M5 brane BPS domain wall in the 3d $\mathcal{N}=2$ theory obtained by compactifying M-theory on the Calabi-Yau fourfold as described above. To this end, we first need to identify the corresponding 3d theory. The bosonic action of eleven dimensional M-theory in the supergravity limit contains the Chern-Simons interaction
\begin{equation} \label{eq:11d}
    S_{11d} \sim \int_{M_{11}} C \wedge G \wedge G,
\end{equation}
where $C$ is the M-theory 3-form and $G$ its field strength, $G \sim dC$. Next, we want to compactify this action along the fourfold. Since we are looking for domain wall solutions arising from M5 branes wrapped on the 4-cycle $P$, we must pick a 4-form flux for $G$ which jumps when crossing the domain wall \cite{Gukov:1999ya}. Flux quantization in M-theory requires that the cohomology class of $G/2\pi$ is a characteristic class given by \cite{Witten:1996md}
\begin{equation}
    \left[\frac{G}{2\pi}\right] = \xi \in H^4(X,\mathbb{Z}) + c_2(X)/2,
\end{equation}
where in our case, since $X$ is non-compact, we get
\begin{equation}
    c_2(X) = c_2(T^*P) = -c_1(P)^2 + 2 c_2(P).
\end{equation}
In the case of $P= \mathbb{P}^2$, for example, one thus gets $c_2(X) = 3 H \wedge H$, where $H$ is the hyperplane class of $\mathbb{P}^2$. Now, when $N$ M5 branes are wrapping $P$, on one side of the domain wall we will have the characteristic class 
\begin{equation}
    \xi_1 = N [P] + c_2(X)/2,
\end{equation}
with $N$ being the number of fivebranes, while on the other side the condition is
\begin{equation}
    \xi_2 = c_2(X)/2,
\end{equation}
which together guarantee that $\xi_1 - \xi_2 = N [P]$. Next, we are ready to perform the compactification on the fourfold. In order to get a sensible result, we can first blow up the ADE singularity along the fiber direction and expand
\begin{equation}
    C = \sum_{i} a_i \wedge B_i + b_i \wedge H_i,  \quad B_i, H_i \in H^2(X,\mathbb{Z}),
\end{equation}
where $B_i$ are two-forms which are Poincare dual to blow-up cycles of the resolved singularity and the $H_i$ span the second cohomology of $P$. Moreover, the $a_i$ and $b_i$ are one-forms with support on the remaining $\mathbb{R}^3$ perpendicular to the Calabi-Yau. For the 4-forms $G_i$ ($i=1,2$) on the two sides of the domain wall we then obtain the condition
\begin{eqnarray}
    G_1 & = & \sum_i da_i \wedge B_i + db_i \wedge H_i + N [P] + c_2(X)/2, \\
    G_2 & = & \sum_i da_i \wedge B_i + db_i \wedge H_i + c_2(X)/2.
\end{eqnarray}
From now on, for the sake of simplicity, we will specialize to the case of $N=1$ and transverse $\mathbb{C}^2/\mathbb{Z}_k$ singularity in the Calabi-Yau. Plugging the above expansions for $C$ and $G$ into equation \eqref{eq:11d}, we compute the following effective 3d actions on the two sides of the domain wall,
\begin{eqnarray}
    S^1_{3d} & \sim & \frac{1}{2}\sum_{i,j} K_{ij} a_i \wedge da_j + \sum_{i,j} Q_{ij} b_i \wedge db_j, \\ 
    S^2_{3d} & \sim & ~ \frac{1}{2} \sum_{i,j} K_{ij} a_i \wedge da_j,
\end{eqnarray}
where $K_{ij}$ denotes the intersection form of the transverse singularity while $Q_{ij}$ is the intersection form of the second cohomology on $P$, 
\begin{equation}
    K_{ij} \equiv B_i \cdot B_j, \quad Q_{ij} \equiv H_i \cdot H_j.
\end{equation}
In our 3d $\mathcal{N}=2$ supersymmetric theory, the terms $\sum_{i,j} K_{ij} a_i da_j$ can be viewed as arising from the Coulomb branch of an $SU(k)$ Chern-Simons theory at level $1$. In fact, this is the expected result in the singular limit of the Calabi-Yau fiber. The number of vacua of such a theory, both on the Coulomb branch and in the non-Abelian phase, is known to be $k$. This can be seen for example as follows \cite{Gadde:2013sca}. Upon compactification on a circle, the 3d theory becomes a 2d $\mathcal{N}=(2,2)$ theory with twisted superpotential given by 
\begin{equation}
    \widetilde{W} = \sum_{i,j} \frac{K_{ij}}{2} \log x_i \cdot \log x_j,
\end{equation}    
and dynamical fields $\sigma_i = \log x_i$. Extremizing this superpotential with respect to the dynamical fields $\sigma_i$ gives the equations for supersymmetric vacua
\begin{equation}
    \exp\left(\frac{\partial \widetilde{W}}{\partial \log x_i}\right) = 1.
\end{equation}
For $K_{ij}$ being the Cartan matrix of $SU(k)$, there are exactly $k$ solutions to these equations. On the other side of the domain wall we have two Chern-Simons theories, one with again $k$ vacua and the other, with level matrix $Q_{ij}$, giving rise to $\sigma \equiv \mathrm{sign}(Q)$ degrees of freedom\footnote{In case the intersection form has both positive and negative eigenvalues, the number of vacua is determined by the gravitational anomaly which is equal to the signature of $Q_{ij}$, see for example \cite{Delmastro:2019vnj}. This can be seen by noting that positive values contribute to the left central charge of the boundary CFT and negative values to the right-moving degrees of freedom and only the difference is effective.}. Here we understand $\mathrm{sign}(Q)$ to be the signature of a matrix $Q$. We thus see that the total number of domain walls is
\begin{equation}
    \#(\mathrm{Domain~Walls}) = k^2 \sigma.
\end{equation}
If we assume that each domain wall contributes $1/8$\footnote{Note that the topological central charge $\sigma$ is only well-defined mod $8$.} to the total left-moving central charge, this result matches the value for $c_L$ obtained from the reduction of the 6d anomaly polynomial along the fourmanifold, see Table \ref{tab:6d2dcc}. For the $SU(k)$ theory, that central charge is 
\begin{equation}
    c_L = \frac{1}{4}(\chi - \sigma) + \frac{k^2 \sigma}{8},
\end{equation}
where we will later argue that the term $\frac{1}{4}(\chi - \sigma)$ comes from the reduction of the degrees of freedom associated to the 6d tensor multiplet. 

\section{$\cN=(1,0)$ theory on K\"ahler manifold with MSW twist}
\label{sec:MSWtwist}

In this section, we will compute the dimensional reduction of the anomaly polynomials of 6d SCFTs over 4-manifolds without boundary. 
This will give the anomaly polynomials of 2d SCFTs obtained from such a compactification. Among the information we extract are the central charges of the resulting 2d conformal field theories.

\subsection{Anomaly polynomials in 6D}

We will review how to compute the anomaly polynomials of various 6d SCFTs. There are two types of SCFTs in six dimension, the $\cN=(2,0)$ theories and the more extended class of $\cN=(1,0)$ theories \cite{Heckman:2013pva}.
We will consider the anomaly polynomials for both of these in the following.

\subsubsection{Anomaly polynomials of $\cN=(2,0)$ SCFTs}

The $\cN=(2,0)$ SCFTs in 6d have an ADE classification which enables a concise expression of the corresponding anomaly polynomials for all such theories. 
 Let $G=A_n,D_n,E_n$ denote the ADE type of the theory. Then, the anomaly eight-form \cite{Ohmori:2014kda} is 
\begin{equation} \label{eqn:anomaly-20}
I_8[G]= r_G I_8(1) + d_G h_G \frac{p_2(NW)}{24}.
\end{equation}
In the above expression,
\begin{equation*}
I_8(1)=\frac{1}{48}\left[ p_2(NW)-p_2(TW)+\frac14\bigl(p_1(TW)-p_1(NW)\bigr)^2\right] \;,
\end{equation*}
is the anomaly polynomial for one M5-brane, $NW$ and $TW$ are the normal and tangent bundles of the worldvolume denoted by $W$, respectively, and $r_G$, $d_G$ and $h_G$ are the rank, the dimension, and the dual Coxeter number of the Lie algebra of type $G$.

%
%

\subsubsection{Anomaly polynomials of $\cN=(1,0)$ SCFTs}

Compared with the $\cN=(2,0)$ case, the classification of 6d $\cN=(1,0)$ theories is much more involved. 
When it comes to anomaly polynomials, there does not exist a general formula for all such theories and one needs to work out the corresponding expressions on a case by case basis. 
Here, we will follow \cite{Ohmori:2014kda} to review the basic steps to compute the anomaly polynomials for 6d $\cN=(1,0)$ SCFTs.

The $6$d SCFTs are strongly coupled theories in the UV, and thus a direct computation of the anomaly polynomial is not possible. 
To begin with, one needs to consider the tensor branch of this theory where there exists a Lagrangian description. 
There are three types of $\cN=(1,0)$ multiplets, tensor, vector and hyper multiplets. 
For tensor branch theories without gauge fields, for example E-string theories, one can obtain the anomaly polynomial from the anomaly inflow \cite{Ohmori:2014pca} of M5 branes in M-theory or from the Chern-Simons terms \cite{Ohmori:2014kda} of the corresponding 5D theories after the compactification on a circle.

The tensor branch theory for the more general $\cN=(1,0)$ theories contains the contributions of the vector multiplets. For theories describing $N$ full M5-branes on the ALE singularity $\bC^2/\Gamma$, 
the tensor branch theories include $N-1$ free tensor multiplets, describing the relative positions of the M5-branes and a linear quiver gauge theory $[G_0]\times G_1\times \cdots \times G_{N-1} \times [G_N]$ with $(N-1)$ gauge factors $G_{1,\ldots,N-1}$ and flavor symmetry $G_0\times G_N$. The bifundamental matter charged under $G_i\times G_{i+1}$ describing a single M5 brane probing $\Gamma$ singularity is called ``conformal matter''. 
Depending on the details of the particular 6d theory, the one-loop anomaly polynomial $I^{\mathrm{one-loop}}$ can be expressed in terms of the anomaly polynomial of each such multiplet. We collect the results for the individual multiplets in the Appendix A and for conformal matter see below. 
The one-loop anomaly is given by
\begin{equation}
I^{\text{one-loop}}=\sum_{i=0}^{N-1} I^\text{bif}_{G,G}(F_i,F_{i+1})+\sum_{i=1}^{N-1} I^\text{vec}_G(F_i) +NI^\text{tensor} .
\end{equation}
Here, we include the center of mass tensor multiplet for convenience.

The resulting expression for the one-loop anomaly polynomial contains contributions of gauge anomalies, mixed gauge and R-symmetry anomalies, mixed gauge and flavor anomalies, as well as mixed gauge and gravitational anomalies. 
Let $n_T$ be the total number of tensor multiplets and $\Omega^{ij}$ be the associated charge lattice. 
One can modify the Bianchi identity of the self-dual two-forms in each of these $n_T$ tensor multiplets by 
\begin{equation}
d\mathcal{H}_i = {\cal I}_i =  \frac{1}{4} \Tr F_i^2- \frac{1}{4} \Tr F_{i+1}^2 + \frac{1}{2}(2i- N+1) |\Gamma | c_2(R),   \label{eq:calH}
\end{equation}
with $i=1,2,\ldots,n_T$ such that the Green-Schwarz contribution $I^{\text{GS}}=\frac12 \sum_{i=0}^{N-1} I_i I_i$
can exactly cancel the above mentioned pure and mixed gauge anomalies in $I^{\mathrm{one-loop}}$.

To obtain the anomaly polynomial of the SCFT, one needs to subtract the contribution from the center of mass tensor multiplet, which is given by 
\begin{equation}
I_{8}^{\mathrm{center-of-mass}}= I_8^{\text{ten}}-\frac1{2N}\left( \frac14 \Tr F^2_0 -\frac14 \Tr F_N^2\right)^2,
\end{equation}
where the last term accounts for the subtraction of the center of mass term. 
The final result \cite{Gukov:2018iiq} is 
\begin{align} \label{eqn:anomaly-10}
    I_{8}^{\mathrm{SCFT}} 
    &= I_{8}^{\mathrm{one-loop}} +  I_{8}^{\mathrm{GS}} - I_{8}^{\mathrm{center-of-mass}} \nonumber \\
    &=\alpha c_2(R)^2+\beta c_2(R)p_1(T)+\gamma p_1(T)^2 +\delta p_2(T) \nonumber \\
    &\qquad +\sum_i^{n_F} \left(\epsilon_i c_2(R) + \zeta_i p_1(T)\right) \tr{F_i^2} + I_8(F^4),
\end{align}
where $n_F$ is the number of the flavor symmetries, $\alpha, \beta, \gamma, \delta, \epsilon_i, \zeta_i$ with $i=1,2,\ldots,n_F$ are rational numbers depending on the quiver structure and $I_8(F^4)$ denotes the terms quartic in the field strength of the background flavor fields.  
This approach can calculate the anomaly polynomials of any $\cN=(1,0)$ theories containing vector multiplets. 
We will see an example in the following.

\paragraph{Simple conformal matter.}

For a single M5-brane probing an ADE singularity, we will get ADE-type conformal matter theories, whose anomaly polynomials have been computed in \cite{Ohmori:2014kda}. 
We sum up their results below \cite{Ohmori:2014kda, Heckman:2018jxk}:
\begin{align}
\begin{split}
    I_{G,G}(F_L,F_R)&=\frac{a}{24}c_2(R)^2-\frac{b}{48}c_2(R)p_1(T)+c\frac{7p_1(T)^2-4p_2(T)}{5760} \\
    & \quad +\left(-\frac{x}{8}c_2(R)+\frac{y}{96}p_1(T)\right)\left(\Tr F_L^2+\Tr F_R^2\right)+\frac{t}{768}\left(\Tr F_L^4+\Tr F_R^4\right) \\
    & \quad +\frac{z}{32}\left((\Tr F_L^2)^2+(\Tr F_R^2)^2 \right) + \frac{w}{16} \Tr F_L^2 \Tr F_R^2 ,
\end{split}
\end{align}
where $G$ spedifies the ADE-type of the singularity, $F_L$ and $F_R$ are the field strengths of the flavor symmetries of the conformal matters, and the coefficients of $a$, $b$, $c$, $x$, $y$, $t$, $z$ and $w$ are group theoretical data summarized in Table \ref{tab:6dcc}.
\begin{table}
\centering
\begin{tabular}{|c||c|c|c|c|c|}
\hline
$G$ & ${\rm SU}(k)$ & ${\rm SO}(2k)$ & $E_6$ & $E_7$ & $E_8$ \\ \hline\hline
$a$ &  0 &  $10k^2-57k+81$ & 319 & 1670 & 12489\\ \hline
$b$ & 0 & $2k^2-3k-9$ & 89 & 250 & 831\\ \hline
$c$ & $k^2$ & $2k^2-k+1$ & 79 & 134 & 249 \\\hline 
$x$ & 0 & $2k-6$ & 12 & 30 & 90 \\\hline
$y$ & $k$ & $2k-2$ & 12 & 18 & 30 \\\hline 
$t$ & $k$ & $k-4$ & 0 & 0 & 0 \\\hline 
$z$ & $0$ & $1$ & 2 & 3 & 5 \\\hline 
$w$ & $1$ & $1$ & 1 & 1 & 1 \\\hline 
\end{tabular}
\caption{Parametrization for anomaly polynomials of 6d conformal matter theories of ADE type.}
\label{tab:6dcc}
\end{table}
Notice that when $G$ is of A type, the ``conformal matter'' is a Lagrangian hypermultiplet bifundamental in $SU(k)\times SU(k)$. To obtain the anomaly polynomial of a single M5 brane probing a $\bZ_k$ singularity, one needs to add the contribution of a tensor multiplet. Thus, the anomaly polynomial is
\begin{align}
\begin{split}
    I_8 &=
    \frac{c^2_2(R)}{24}+\frac{c_2(R)p_1(T)}{48}
    +\frac{7k^2+23}{5760}p^2_1(T)
    -\frac{k^2+29}{1440}p_2(T) \\
    & \qquad
    +\frac{k(\Tr F^2_L+ \Tr F^2_R)}{96} p_1(T)
    +\frac{k(\Tr F^4_L+\Tr F^4_R)}{768}
    +\frac{\Tr F^2_L \Tr F^2_R}{16}.
\end{split}
\end{align}
%

\paragraph{Class $S_k$.}

\begin{figure}[ht]
    \centering
    
    \resizebox{.4\textwidth}{!}{%
    \begin{tikzpicture}
        \tikzstyle{gauge} = [circle,inner sep=6pt, draw];
	    \tikzstyle{flavour} = [regular polygon,regular polygon sides=4,inner sep=6pt, draw];
        \node (f0) [flavour,              label=above:{$\scriptscriptstyle{SU(k)}$}] {};
        \node (g1) [gauge,   right of=f0, label=above:{$\scriptscriptstyle{SU(k)}$}] {};
        \node (l1) [         right of=g1] {$\ldots\ldots$};
        \node (g2) [gauge,   right of=l1, label=above:{$\scriptscriptstyle{SU(k)}$}] {};
        \node (f3) [flavour, right of=g2, label=above:{$\scriptscriptstyle{SU(k)}$}] {};
        \draw (f0) --(g1) (g1)--(l1) (l1)--(g2) (g2)--(f3);
        \draw[decoration={brace,raise=10pt},decorate,thick]
         (3.25,-0.1) -- node[below=10pt] {$\scriptstyle{N-1}$} (0.75,-0.1);
    \end{tikzpicture}
    }
    \caption{The $S_k$ class in tensor branch}
    \label{fig:skQ}
\end{figure}
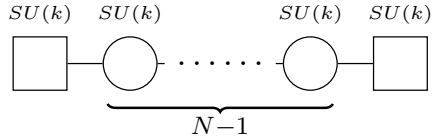

Consider the $\cN=(1,0)$ theories of $N>1$ M5 branes probing a $\bC^2/\bZ_k$ singularity. 
The tensor branch is described by a linear quiver diagram depicted in Figure \ref{fig:skQ}. 
One can find the following $\cN=(1,0)$ multiplets on the tensor branch:
\begin{itemize}
    \item $n_T = N-1$ tensor multiplets,
    \item $n_V = N-1$ ($n_F=2$) vector multiplets with gauge (flavor) group $SU(k)$,
    \item $n_H=N$ hyper multiplets in bi-fundamental representation of $\left[SU(k) \times SU(k)\right]$.
\end{itemize}

Let $F_i$ be the field strength associated with the gauge nodes $i=1,\ldots,N-1$ and flavor node ($i=0$ and $i=N$) in Figure \ref{fig:skQ}. The one-loop anomaly polynomial is 
\begin{align}
    I^{\mathrm{one-loop}} & = \sum_{i=0}^{N-1} I_{8}^{\mathrm{hyper}}(F_i,F_{i+1}) + \sum_{i=1}^{N-1} I_{8}^{\mathrm{vector}}(F_i) + (N-1)I_{8}^{\mathrm{tensor}}(F) .
\end{align}
Now, let's focus on the part containing the gauge anomalies,
\begin{equation} \label{SKgauge}
    I^{\mathrm{one-loop}} \supset -\frac{1}{16}\sum_{i=1}^{N-1}(\Tr F_i^2)^2 +\frac{1}{16} \sum_{i=0}^{N-1}\Tr F_i^2 \Tr F_{i+1}^2 - \frac{k}{4}c_2(R) \sum_{i=1}^{N-1}\Tr F_i^2~.
\end{equation}
Let $H_i$ be the field strength of the two-form in the $i$th tensor multiplet. One can modify the Bianchi identity to be $dH_i=I_i$ in such a way that all the gauge dependent anomalies in equation \eqref{SKgauge} are canceled. 
In this example, the $I_i$ are determined to be 
\begin{align} \label{eqn:GS_sk}
	I^i=\Omega^{ij}I_j=\frac{1}{4}(2\Tr F_i^2-\Tr F_{i-1}^2 -\Tr F_{i+1}^2 )+k c_2(R),
\end{align}
where $\Omega^{ij}$ is the intersection form on the charge lattice 
 \begin{equation}
\Omega^{ij}=\begin{pmatrix}
2 &- 1 & \\
-1 & 2 & -1 \\ 
& \ddots & \ddots & \ddots \\
&&-1&2&-1\\
&&&-1 & 2
\end{pmatrix}.
\end{equation}
Taking into account the Green-Schwarz contribution specified in equation (\ref{eqn:GS_sk}),
one then arrives at the final result,  
\begin{align} \label{eqn:6danomalySK}
\begin{split}
I^{\text{scft}}_8 =& \frac{c_2(R)^2}{24} \left[k^2N^3-2(k^2-1)N+K^2-2\right]
-\frac{1}{48}(N-1)(k^2-2) c_2(R)p_1(T) \\
&+\frac{k}{24}(\Tr F_0^4+\Tr F_N^4)
+\frac{30 N+7k^2-30}{5760}p_1(T)^2 
-\frac{30 N+k^2-30}{1440}p_2(T) \\
&-\frac{k(N-1)}{8}c_2(R)(\Tr F_0^2+\Tr F_N^2)  
+\frac{k}{96}p_1(T)(\Tr F_0^2+\Tr F_N^2) \\
&+\frac{1}{32}((\Tr F_0^2)^2+(\Tr F_N^2)^2)
-\frac{1}{32N}(\Tr F_0^2+\Tr F_N^2)^2.  
\end{split}
\end{align}

\subsection{Anomaly polynomial reduction on K\"ahler surfaces with MSW twist}
\label{sec:MSWtwist2}

We will study the dimensional reduction of anomaly polynomials in the compactification of 6d SCFTs over K\"ahler 4-manifolds $M_4$. 
We will consider both the $\cN=(2,0)$ and $\cN=(1,0)$ SCFTs. The $6$d theories are put on the geometry $\Sigma \times M_4$ where $\Sigma$ is a Riemann surface and $M_4$ is a K\"ahler 4-manifold. Moreover, we assume that both $\Sigma$ and $M_4$ are Euclidean. 
To preserve supersymmetry in the effective theory, one needs to perform a topological twist. 
The anomaly polynomial in the 2d effective theory is a 4-form $I_4$. 
It can be obtained by integrating the degree-8 anomaly polynomial $I_8$ of the 6d theory over $M_4$.
As we will see later in this section, one can obtain the central charge of the  effective theory from the anomaly polynomial $I_4$.

\subsubsection{Reduction of anomaly polynomials for $\cN=(2,0)$ SCFTs}
\label{sec:MSWa}

First, let's consider an $\cN=(2,0)$ SCFT on $\Sigma \times M_4$. The supercharges of the theory transform as $\bf{(4^+,4)}$ under $SO(6)\times SO(5)_R$. Since $M_4$ is K\"ahler, the holonomy group is reduced to $U(2)$.
The Lorentz group and R-symmetry group decompose as 
\begin{eqnarray*}
SO(6)   
\quad  \to &   \quad 
SU(2)_l \times SU(2)_r \times U(1)_{\Sigma} 
& \to  \quad 
SU(2)_l \times U(1)_r \times U(1)_{\Sigma} ,
\nonumber \\
\bf{4^+} 
\quad  \to &  
\bf{(2,1)}_1 + \bf{(1,2)}_{-1} 
& \to  \quad 
\bf{2}_{0,1} + \bf{1}_{\pm 1,-1} 
\nonumber \;,
\end{eqnarray*}
and
\begin{eqnarray}
& SO(5)_R & \to \quad SU(2)_R \times U(1)_t, \nonumber \\
& {\bf 4} & \to \quad {\bf 2}_{\pm 1}.  \nonumber
\end{eqnarray}
Then, after performing the twist $U(1)_{\text{tw}} = U(1)_r \times U(1)_t$, the representations transform as 
\begin{eqnarray}
    SO(6)  \times SO(5)_R & \to  & \quad SU(2)_R \times SU(2)_l \times U(1)_{\text{tw}} \times U(1)_{\Sigma}, \nonumber \\
      \bf{(4^+,4)}  & \to & \quad \bf{(2,2)}_{\pm 1 ,1} + \bf{(2,1)}_{\pm 2,-1} + \bf{(2,1)}_{0,-1}+
      \bf{(2,1)}_{0,-1}.
\end{eqnarray}
The two $\bf{(2,1)}_{0,-1}$ occurrences are singlets under $SU(2)_l \times U(1)_{\text{tw}}$ and doublets under the R-symmetry $SU(2)_R$.
Thus, after compactification, one should have a 2d effective theory with supersymmetry $\cN=(0,4)$ which is the expected amount of supersymmetry for M5 branes wrapping a K\"ahler 4-cycle in a Calabi-Yau threefold, giving rise to the MSW CFT. 
Equivalently, the above result can be also obtained by first performing a Vafa-Witten twist along a general $M_4$ by $SU(2)_{\text{tw}}=\text{Diag}[SU(2)_r \times SU(2)_R]$ and then considering the following decomposition $SU(2)_{\text{tw}} \to U(1)_{\text{tw}}$ when $M_4$ is K\"ahler \cite{Apruzzi:2016nfr, Gukov:2018iiq}.

Let's consider the dimensional reduction of the anomaly polynomial for the MSW twist. In the compactification, the Pontryagin classes for the tangent bundle $TW$ and the normal bundle $NW$ decompose as 
\begin{eqnarray*}
& p_1(TW) = p_1(T\Sigma) + p_1(TM_4), \qquad & p_1(NW) = p_1(R) + p_1(t),\\
    & p_2(TW) =  p_1(TM_4)p_1(T\Sigma), \qquad &p_2(NW) = p_1(R)p_1(t),
\end{eqnarray*}
where $T\Sigma$ and $TM_4$ denote the tangent bundles of $\Sigma$ and $M_4$, respectively, and $R$ and $t$ denote the bundle corresponding to the $SU(2)_R$-symmetries and $U(1)_t$-symmetries.
Here, the 6d R-symmetry is $SO(5)_R \subset SU(2)_R \times U(1)_t$. 
The topological twist is realized by substituting $c_1(t) \to c_1(t)+c_1(M_4)$, where we refer to \cite{Apruzzi:2016nfr} for more details. 
Using the fact that $p_1(t)=c_1(t)^2$ and $\int_{M_4}c_1^2(M_4)=2\chi+3\sigma$, we perform the integral of the anomaly polynomial $I_8$ over $M_4$, giving
\begin{equation} \label{eq:I8M5M4}
    \int_{M_4} I_8 = \frac{r_G}{48}\left[
    -(\chi+3\sigma)p_1(T\Sigma)+3(\chi+\sigma)p_1(R)\right]+d_G h_G \frac{2\chi+3\sigma}{24}p_1(R).
\end{equation}

The anomaly polynomial of general 2d $\cN=(0,4)$ theories has the following form \cite{Harvey:1998bx},
\begin{equation}
    I_4 = \frac{c_L-c_R}{24} p_1(T\Sigma)+\frac{c_R}{24}p_1(R),
\end{equation}
where $p_1(R)$ is the first Pontryagian class of the $SU(2)_R$ bundle. Comparing with \eqref{eq:I8M5M4}, we find  
\begin{align} \label{eqn:c20}
    c_R &= \frac{3}{2}(\chi+\sigma)r_G+(2\chi+3\sigma)d_G h_G, \nonumber \\
    c_L &= \chi r_G+(2\chi+3\sigma)d_G h_G, 
\end{align}
which are the same as the central charges obtained by the Vafa-Witten twist in \cite{Alday:2009qq}.
In particular, for a single M5 brane, the 2d central charges are
\begin{equation} 
   c_L = \chi, \qquad c_R =\frac{3}{2}(\chi+\sigma),
\end{equation}
which reproduce the well-known central charges of the MSW CFT.

\paragraph{MSW CFT.}

Consider the configuration of a single M5 brane wrapping a K\"ahler four-cycle $P$ inside a Calabi-Yau threefold. The IR limit of the 2d effective theory is believed to be an $\cN=(0,4)$ SCFT. 
Here, the right-moving chiral algebra is the ``small" $\cN=4$ superconformal algebra with R-symmetry $SU(2)_R$. 
By dimensional reduction of a free 6d $\cN=(2,0)$ tensor multiplet and counting of possible 2d massless fields, 
one can obtain the following central charges \cite{Maldacena_1997}
\begin{align} \label{eqn:CCmsw}
    &c_L = 2h^{2,0} + h^{1,1}+2+2h^{0,1} = \chi, \nonumber \\
    &c_R = \frac{3}{2}(4h^{2,0}+4)=\frac{3}{2}(\chi + \sigma),
\end{align}
where we have used the fact that $b^+_2 = 2h^{2,0}+1$ and $b_2^-=h^{1,1}-1$ for K\"ahler surfaces. 
Here, we also assume that $b_1(P)=0$. 
The above result derived by counting massless fields matches the anomaly inflow computation \cite{Harvey:1998bx}. 
In addition, the number of the right-moving bosonic degrees of freedom is 
a multiple of four as for
a non-linear sigma model with $\cN=4$ supersymmetry, the bosons should span a hyperk\"ahler manifold whose real dimension is divisible by four.
The R-symmetry of the small $\cN=4$ superconformal algebra is affine $SU(2)_k$ with the central charge $c_R=6k$. From the result above, one can read off the level to be $k=(\chi+\sigma)/4 = h^{2,0}+1$, which is an integer as expected. 
However, for $b_1(P)\neq 0$, there is a mismatch due to some of the massless fields becoming massive along the RG flow. 

\subsubsection*{Central charge from the reduction of a single M5 brane}

The worldvolume theory of a single M5 brane is a 6d Abelian $(2,0)$ SCFT. 
There are 16 supercharges organized as 4 symplectic Majorana-Weyl spinors transforming as $\bf{4}$ under the R-symmetry $SO(5)_R$. 
The field content of this theory is just a free 6d $(2,0)$ tensor multiplet made up of one $\cN=(1,0)$ tensor multiplet and one $\cN=(1,0)$ hypermultiplet. It contains a self-dual 2-form $B^+_{MN}$, two complex chirality $+$ spinors $\psi^+$ and 5 scalar $t_I$ with $I=0,1,\ldots,4$ transforming as $\bf{1}$, $\bf{4}$ and $\bf{5}$ under $SO(5)_R$.

After the MSW twist along the K\"ahler manifold $M_4$, the twisted 6d fields transform as
\begin{eqnarray}
& SO(6) \times SO(5)_R & \to \quad 
SU(2)_R  \times SU(2)_l \times U(1)_{tw} \times U(1)_{\Sigma} ,
\nonumber \\
& B^+_{MN} =\bf{(15^+,1)} & \to \quad  \bf{(1,1)}_{0,0} + \bf{(1,2)}_{\pm 1, \pm 2} +  \bf{(1,3)}_{0,0} + \bf{(1,1)}_{0,0} + \bf{(1,1)}_{\pm 2,0} ,\nonumber\\
& H^+_{MNL} = \bf{(10^+,1)} & \to \quad  \bf{(1,2)}_{\pm 1,0} + \bf{(1,3)}_{0, 2} +  \bf{(1,1)}_{0,-2} + \bf{(1,1)}_{\pm 2,-2}, \nonumber\\
& t_i=\bf{(1,5)} & \to \quad {\bf (1,1)}_{\pm 2,0} + {\bf (3,1)}_{0,0}, \nonumber\\
& \psi^+= \bf{(4^+,4)} & \to \quad 
\bf{(2,2)}_{\pm 1, 1}
+
\bf{(2,1)}_{0, -1}
+
\bf{(2,1)}_{0, -1}
+
\bf{(2,1)}_{\pm 2,-1}.
\end{eqnarray}
After reduction along $M_4$, we thus obtain the following field content in two dimensions:
\begin{itemize}
    \item The contribution of the self-dual two-form $B_{MN}$ is counted in terms of the three-form $H^+_{MNL}$. After the dimensional reduction, $\bf{(1,3)}_{0, 2}$ contributes $b^-_2$ left-moving scalars, $\bf{(1,1)}_{0,-2}$ contributes one right-moving scalar and $\bf{(1,1)}_{\pm 2,-2}$ contributes $2h^{2,0}$ right-moving scalars. Since $b_2^+=2h^{2,0}+1$ for K\"ahler surfaces, there are in total $b_2^-$ left-moving and $b_2^+$ right-moving scalars. 
    
    \item 
    Dimensional reduction of the twisted fields contribute $3$ scalars from ${\bf (3,1)}_{0,0}$, which correspond to the $3$ transverse directions of the M5 branes inside $\bR^5$ after compactification on the $CY_3$ manifold.
    There are also $2h^{2,0}$ scalars from ${\bf (1,1)}_{\pm 2,0}$ corresponding to the holomorphic moduli of the K\"ahler cycle inside the $CY_3$ manifold. 
    In total, we have $2+b_2^+$ scalars. 
    
    \item 
     Dimensional reduction of the 2 complex spinor $\psi^+$ after the topological twist contribute 4 right-moving spinors from $\bf{(2,1)}_{0, -1}$ and $4h^{2,0}$ right-moving spinors from $\bf{(2,1)}_{\pm 2,-1}$. In total, there are $2+2b_2^+$ right-moving spinors. 
\end{itemize}
\begin{table}[h!]
\centering
\begin{tabular}{|c|c|c|} 
\hline
6d fields &     Left     &     Right     \\ \hline
$B^{+}_{MN}$ &  $b_2^-$ compact bosons & $b_2^+$ compact bosons \\ \hline
$t_I$ &  $b_2^+ + 2$  non-compact bosons & $b_2^+ + 2$  non-compact bosons \\ \hline
$\psi^+$ & & $2(b^+_2+1)$ real fermion\\
\hline
\end{tabular}
\caption{Reduction of the 6d $(2,0)$ tensor multiplet along a K\"ahler 4-manifold.}
\label{tab:20tensorred}
\end{table}

The field content after the compactification is summarised in the table above. 
Taking all these fields into account, the left and right moving central charges are
\begin{align}
\begin{split}
    &c_L \; = \; (2b_0 + b_2^+) +b_2^- = \chi, \\
    &c_R = (2b_0 + b_2^+) + b_2^+ + \frac{1}{2}(2b_0+2b_2^+) =\frac{3}{2}(\chi+\sigma), 
\end{split}
\end{align}
which agrees with the result obtained from the dimensional reduction of the anomaly polynomial.

\subsubsection{Reduction of anomaly polynomials for $\cN=(1,0)$ SCFTs}
\label{sec:MSWb}

Let us now consider the $\cN=(1,0)$ theories on $\Sigma \times M_4$.  
Similarly to the $\cN=(2,0)$ theories, the 2d effective theory after an MSW twist has $\cN=(0,2)$ supersymmetry \cite{Gukov:2018iiq}. 
Considering the twist $U(1)_{\text{tw}}=U(1)_r \times U(1)_t$ where $U(1)_t$ is a subgroup of $SU(2)$, the supercharges transform as 
\begin{eqnarray}
    SO(6)  \times SU(2)_R 
    & \to  & \quad 
    SU(2)_l \times U(1)_{\text{tw}} \times U(1)_{\Sigma} 
    \nonumber \\
     \bf{(4^+,2)}  
     & \to & \quad 
\bf{1_{0,-1}} 
+
\bf{1_{0,-1}}
+
\bf{1_{\pm 2, -1}}
+
\bf{2_{\pm 1, 1}}
\;.
\end{eqnarray}
Both of the two supercharges in the $\bf{ 1_{0,-1}}$ representation can be made covariantly constant along $M_4$ and the effective 2d theory will have $(0,2)$ supersymmetry. 
Analogous to the 6d $\cN=(2,0)$ case, the same result can be derived by first performing a Vafa-Witten twist $SU(2)_{\text{tw}}=\text{Diag}[SU(2)_r \times SU(2)_R]$ for a general 4-manifold $M_4$ and subsequently decomposing under $SU(2)_{\text{tw}} \supset U(1)_{\text{tw}}$ when $M_4$ is K\"ahler \cite{Apruzzi:2016nfr, Gukov:2018iiq}.


The anomaly polynomial of the effective 2d theory can be derived by integrating the 8-form $I_8$ defined in equation (\ref{eqn:anomaly-10}) over a 4-manifold. 
Similar to the discussion of $\cN=(2,0)$ theories, to perform this integration, we first implement the following decomposition for the tangent bundle on the worldvolume of the M5 brane, denoted as $TW$,
\begin{eqnarray*}
p_1(TW) = p_1(T\Sigma) + p_1(TM_4), \qquad  
p_2(TW) =  p_1(TM_4) p_1(T\Sigma).
\end{eqnarray*}
We will identify the Cartan subalgebra $U(1)_r \subset SU(2)_R$ as the R-symmetry for the 2d $\cN=(0,2)$ theories.  
Let $c_1(r)$ be the Chern root of the $U(1)_r$ bundle. 
After the topological twist, it is shifted to be 
\begin{equation*}
c_1(r) \to c_1(r) +  \frac{c_1(TM_4)}{2}.
\end{equation*}
The second Chern class thus decomposes as 
\begin{equation*}
c_2(R) = -(c_1(r)+  c_1(TM_4)/2)^2.
\end{equation*}

After the integration of the anomaly polynomial for the general 6d $\cN=(1,0)$ theories from equation (\ref{eqn:anomaly-10}), we get  
\begin{align}
    \int_{M_4} I_8 &= \left[(2\gamma+\delta)3\sigma -\frac{1}{4}(2\chi+3\sigma)\beta\right] p_1(T\Sigma)+\left[\frac{3}{2}\alpha(2\chi+3\sigma)-3\sigma \beta\right]c_1^2(r) \nonumber \\
    &+ \sum_i^{n_F} \left(-\frac{\epsilon_i}{2} \chi + 3(\zeta_i-\frac{\epsilon_i}{4}) \sigma \right)  \tr{F_i^2},
\end{align}
where $\alpha$, $\beta$, $\gamma$, $\delta$, $\epsilon_i$, $\zeta_i$ with $i=1,2,\ldots, n_F$ are the coefficients in the anomaly polynomial and $\chi$ and $\sigma$ denote Euler characteristic and signature of $M_4$.

The anomaly polynomial of a 2d $\cN=(0,2)$ theory has the form 
\begin{equation} \label{eqn:I4-02}
    I_4 = \frac{c_L-c_R}{24}p_1(T\Sigma)+\frac{c_R}{6}\,c_1(r)^2 + I_4(F^2),
\end{equation}
where $I_4(F^2)$ denotes terms quartic in the field strength of the flavor symmetries. 
Comparing with equation (\ref{eqn:I4-02}), one can read off both central charges 
\begin{align*} 
c_R&= 9 \cdot  (3 \alpha -2 \beta )\sigma +18 \alpha  \chi, \\
c_L&= 9 \cdot  (3 \alpha -4 \beta +16 \gamma +8 \delta ) \sigma + 6 \cdot (3 \alpha   -2 \beta)  \chi. \nonumber
\end{align*}

In the following we will present several examples.

\paragraph{Simple conformal matter on K\"ahler surfaces.}

Consider the worldvolume theory of a single M5 brane probing an $ADE$ singularity. The anomaly polynomial after the dimensional reduction is given by
\begin{equation} \label{eqn:I4_ade}
    I_4 = \frac{c_L-c_R}{24}P_1(T\Sigma) + \frac{c_R}{6}C^2_1(R) + (\frac{e_f}{16} \chi + \frac{s_f}{32} \sigma) (\tr{F_L^2}+ \tr{F_R^2}),
\end{equation}
where the central charge of the infrared $\cN=(0,2)$ SCFTs are given by 
\begin{equation} \label{eqn:2dcc_ade}
    c_L= \frac{e_{l}}{2}
     \chi + \frac{s_l}{8} \sigma, \qquad c_R = \frac{3e_r}{4} \chi  + \frac{3s_r}{4} \sigma~.
\end{equation}
%
Here, the parameters $e_l,s_l,e_r,s_r,e_f,s_f$ only depend on the conformal matter type and are organized in Table \ref{tab:6d2dcc}.
\begin{table} 
\centering
\begin{tabular}{|c||c|c|c|c|c|}
\hline
$G$ & ${\rm SU}(k)$ & ${\rm SO}(2k)$  & $E_6$ & $E_7$ & $E_8$ \\ \hline\hline
$e_l$ & $1$    & $16 k^2-87 k+117$   & 523   & 2630 & 19149  \\ \hline
$s_l$ & $k^2-4$ & $103 k^2-531 k+675$ & 3484  & 8332 & 117636  \\ \hline
$e_r$ & $1$     & $(k-3)(16k-39)$     & 638   & 1670 & 12489  \\\hline
$s_r$ & $1$     & $(k-3)(10k-27)$     & 1046  & 2630 & 19149  \\\hline 
$e_f$ & $0$     & $2k-6$     & 12  & 18 & 90  \\\hline 
$s_f$ & $k$     &  $8k-20$    & 48  & 57 & 300  \\\hline 
\end{tabular}
\caption{Parametrization of central charges obtained by reducing conformal matter theories of ADE type along 4-manifolds.}
\label{tab:6d2dcc}
\end{table}

\paragraph{$S_k$ class.}

We perform the dimensional reduction of the 6d anomaly polynomial (\ref{eqn:6danomalySK}) on general K\"ahler 4-manifolds. 
Comparing the result with the equation (\ref{eqn:I4-02}), we can extract the left/right moving central charge 
\begin{eqnarray} \label{eqn:2dcc_sk}
c_L&=&   \left(k^2 \left(3 N^3-5 N+2\right)+4 (N-1)\right) \frac{\chi}{4} +\left(k^2 \left(9 N^3-12 N+4\right)-1\right)\frac{\sigma}{8},  \\
 c_R&=& \frac{3}{4} (N-1) \left( 
  \left(k^2 \left(N^2+N-1\right)+2\right)\chi
  +
  \left(k^2 \left(3 N^2+3 N-2\right)+4\right)\frac{\sigma}{2} \right), \nonumber
\end{eqnarray}
and the flavor dependent term  
\begin{equation}\label{eqn:2dF_sk}
    I_4(F^2) = \left(\frac{(N-1)k}{16}\chi +\frac{(3N-2)k}{32}\sigma\right)(\Tr F_0^2+\Tr F_N^2)~.
\end{equation}
Notice that the 2d anomaly polynomial of the dimensional reduction of class $S_k$ can also be rewritten in the form of equation (\ref{eqn:I4_ade}).
It seems that the $\chi$ and $\sigma$ dependence in the 2d anomaly polynomial has the same structure for all $\cN=(1,0)$ theories.

\subsubsection*{Central charge from the dimensional reduction of $\cN=(1,0)$ tensor multiplet}

The 6d $\cN=(1,0)$ theories have eight supercharges with R-symmetry $SU(2)_R$. There are three supermultiplets: the the tensor multiplet, vector multiplet and hypermultiplet.
In specific, the tensor multiplet includes a self-dual 2-form $B^+_{MN}$, a real scalar $t_0$ and a complex Weyl spinor $\psi^+$ transforming as $\bf{2}$ under the $SU(2)_R$ symmetry.

%
    
    
%

After the MSW twist, the fields in the $\cN=(1,0)$ tensor multiplet  transform as 
\begin{eqnarray}
& SO(6) \times SU(2)_R & \to \quad 
SU(2)_l \times U(1)_{tw} \times U(1)_{\Sigma},
\nonumber \\
& B^+_{MN} = \bf{(15^+,1)} & \to \quad  \bf{1}_{0,0} + \bf{2}_{\pm 1, \pm 2} +  \bf{3}_{0,0} + \bf{1}_{0,0} + \bf{1}_{\pm 2,0} ,\nonumber\\
& H^+_{MNL} = \bf{(10^+,1)} & \to \quad  \bf{2}_{\pm 1,0} + \bf{3}_{0, 2} +  \bf{1}_{0,-2} + \bf{1}_{\pm 2,-2} ,\nonumber\\
& t_0=\bf{(1,1)} & \to \quad {\bf 1}_{ 0,0} ,\nonumber\\
& \psi^+= \bf{(4^+,2)} & \to \quad \bf{2}_{\pm 1, 1}
+
\bf{1}_{0, -1}
+
\bf{1}_{0, -1}
+
\bf{1}_{\pm 2, -1}.
\end{eqnarray}

Reduction along $M_4$ then leads to the following field content in two dimensions:
\begin{itemize}

 \item The three-form $H^+_{MNL}$ gives rise to $b_2^-$ left-moving real scalars from $\bf{3}_{0, 2}$, 1 right-moving real scalar from $\bf{1}_{0,-2}$ and $2h^{2,0}$ right-moving real scalars from $\bf{1}_{\pm 2,-2}$. Thus, in total, there are $b_2^+$ right-moving scalars. 

\item The scalar $t_0$ will give $1$ scalar field in 2d effective theory, which corresponds to the transverse direction of the string inside $\bR^3$ after the compactification of M-theory on $CY_4$. Notice that we consider K\"ahler 4-cycles in $CY_4$ which are rigid and without holomorphic deformation. Indeed, for example, if we take $M_4=\bP^2$, then $h^{2,0}=0$. The only non-vanishing Hodge number is $h^{0,0}=h^{1,1}=1$.
In general, one can consider the K\"ahler surfaces with definite negative lattice, i.e. $b_2^+=1$ or $b_2^+=0$ and $b_2^-=h^{1,1}-1>1$.

\item The complex fermions after topological twists will give rise to $2$ right-moving spinors from $\bf{1}_{0, -1}$ and $2h^{2,0}$ right-moving spinors from $\bf{1}_{\pm 2, -1}$. Thus, in total, there are $b_0+b_2^+$ right-moving spinors. 

\end{itemize}

\begin{table}
\centering
\begin{tabular}{|c|c|c|} 
\hline
6d fields &     Left     &     Right     \\ \hline
$B^{+}_{MN}$ &  $b_2^-$ compact bosons & $b_2^+$ compact bosons \\ \hline
$t_I$ &  $1$  non-compact bosons & $1$  non-compact bosons \\ \hline
$\psi^+$ & & $b^+_2+1$ real fermion\\
\hline
\end{tabular}
\caption{Fields obtained from the reduction of the $(1,0)$ tensor multiplet along the 4-manifold.}
\label{tab:tensorreduction}
\end{table}
The results are summarized in the Table \ref{tab:tensorreduction}. From it, the central charges are 
\begin{align*}
    &c_L =  b_0+b_2^- =\frac{1}{2}(\chi-\sigma), \\
    &c_R = (b_0+ b_2^+) + \frac12 (b_0+ b_2^+)=\frac{3}{4}(\chi+\sigma)\;. 
\end{align*}
This is the same as the central charges obtained by the dimensional reduction of the anomaly polynomial for a $\cN=(1,0)$ tensor multiplet 
\begin{equation} I_8^\text{(tensor)}=\frac{c_2(R)^2}{24}+\frac{c_2(R)p_1(T)}{48}+\frac{23p_1(T)^2-116p_2(T)}{5760} \;.
	\label{I-tensor}
\end{equation}

We also studied the dimensional reduction of the free vector-multiplet and hypermultiplets. However, here the central charges obtained by counting the 2d zero modes do not reproduce the central charges obtained by reducing the anomaly polynomial. We leave the investigation of this phenomenon for future work.

\section{Compactification on non-compact 4-manifolds and gluing}
\label{sec:gluing}

In this section, we consider compactifications on non-compact 4-manifolds leading to a coupled 3d-2d system. First, we derive the relevant topological twist to arrive at the relevant 3d theories with a 2d boundaries. Then, we consider \textit{gluing} such 3d-2d systems together by gluing the relevant non-compact four-manifolds along their common boundaries.

\subsection{Compactification on non-compact 4-manifolds and a 3d perspective}

\begin{figure}
    \centering
    \includegraphics{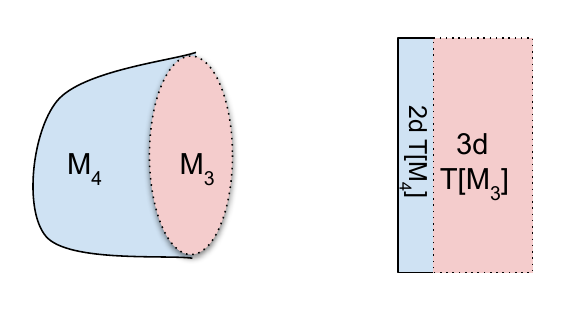}
    \caption{Compactification on a 4-manifold with compact boundary leads to a coupled 3d-2d system.}
    \label{fig:3d2d}
\end{figure}

We begin with compactifications on 4-manifolds bounded by a compact 3-manifold, 
\begin{equation}
    \partial M_4 = M_3,
\end{equation}
where we consider the most general situation such that $M_3$ has $SO(3)$ holonomy. As we will see below, a suitable topological twist along such 3-manifolds upon compactification leads to a 3d $\mathcal{N}=1$ theory in the remaining spacetime dimensions. Now such theories have a mass gap \cite{Agarwal:2012bn} and are expected to flow to TQFTs at low energies. Since we are compactifying on non-compact 4-manifolds, the corresponding 3d TQFT lives on a manifold with boundary and is coupled to a 2d CFT. We propose that this 2d CFT arises from a 2d $\mathcal{N}=(0,2)$ SCFT with a topological twist on the right-moving sector. This coupled 3d-2d system is schematically shown in Figure \ref{fig:3d2d}. If the difference $c_L - c_R$ (modulo $24$) does not vanish, the 3d TQFT requires to choose a well-defined framing on the 3-manifold and is anomalous with the anomaly corresponding to multiplying the amplitudes by integer powers of $\exp\left(2\pi i (c_L-c_R)\right)$ under a change of framing. This is then in turn canceled by a $T$-transformation of the boundary CFT.

\subsubsection*{6d SCFTs on $M_3$ under Vafa-Witten twist}
Consider the 6d $\cN=(1,0)$ theory on $M_3 \times \bR^3$. 
To perform an MSW-like twist, one needs to pick a $U(1)$ subgroup of the holonomy group of $M_3$. There are two situations we'd like to study in detail, namely generic 3-manifolds with $SO(3)$ holonomy and product manifolds of the form $\Sigma \times \mathbb{R}$ where $\Sigma$ is a Riemann surface. The latter case contains a reduced holonomy group $U(1)$ coming from local rotations along the two dimentional subspace $\Sigma$. Let us start by performing the topological twist for general $M_3$ via identifying 
\begin{equation*}
SU(2)_{\text{tw}} = \text{diag}\left[SU(2)_{M_3}\times SU(2)_R \right].
\end{equation*}
The results are 
\begin{itemize}
    \item For $\cN=(2,0)$ theory, the R-symmetry group is $SO(5)_R = SU(2)_R \times U(1)^{\text{3d}}_R$. 
    After twisting, the supercharges transform as 
\begin{eqnarray}
    SO(6)  \times SO(5)_R & \to  & \quad SU(2)_{\text{tw}} \times SU(2)_{\bR^3} \times U(1)^{\text{3d}}_R,
    \nonumber \\
    \bf{(4,4)}  & \to & \quad 
    \bf{(1,2)}_{\pm 1}  + \bf{(3,2)}_{\pm 1}.
\end{eqnarray}
There are four supercharges which leads to a 3d $\cN=2$ theory with a $U(1)_R$ R-symmetry. 
    
    \item For $\cN=(1,0)$ theory, the R-symmetry group is $SU(2)_R$.
    After twisting, the supercharges transform as
    \begin{eqnarray}
    SO(6)  \times SU(2)_R & \to  & \quad  SU(2)_{\text{tw}} \times SU(2)_{\bR^3} ,
    \nonumber \\
    \bf{(4,2)}  & \to & \quad 
    \bf{(1,2)}  +  \bf{(3,2)}.
\end{eqnarray}
There are two supercharges resulting in a 3d $\cN=1$ theory. 
\end{itemize}

\subsubsection*{6d SCFTs on $\Sigma \times \bR$ under MSW twist}
If the metric on $\Sigma$ is chosen to be independent of $S^1$,
the holonomy group reduces from $SO(3)$ to $U(1)_{\Sigma}$ \cite{Gukov:2015sna}. All included, the $SO(6)$ honolomy of the general six manifolds reduce as follows, 
\begin{eqnarray*}
SO(6)   
\quad  \to &   \quad 
SU(2)_{\bR^3}  \times SU(2)_{M_3}
& \to  \quad 
SU(2)_{\bR^3} \times U(1)_{\Sigma} ,
\nonumber \\
\bf{4} 
\quad  \to &  
\bf{(2,2)} 
& \to  \quad 
\bf{2}_{\pm 1} 
\nonumber \;.
\end{eqnarray*}
Performing the MSW twist by
\begin{equation*}
    U(1)_{\text{tw}}=U(1)_{\Sigma} \times U(1)_t
\end{equation*}
where $U(1)_t$ is part of the 6d R-symmetry, one gets
\begin{itemize}
    \item For $\cN=(2,0)$ theory, the R-symmetry group is $SO(5)_R \supset SU(2)_R \times U(1)_t$. 
    After twisting, the supercharges transform as 
\begin{eqnarray}
\begin{split}
    SO(6)  \times SO(5)_R  & \quad \to  \quad SU(2)_R \times SU(2)_{\bR^3} \times U(1)_{\text{tw}}, \\
    \bf{(4^+,4)}  & \quad \to  \quad 
    \bf{(2,2)}_{0}  + \bf{(2,2)}_{ 0}+ \bf{(2,2)}_{\pm 2}.
\end{split}
\end{eqnarray}
There are eight supercharges leading to a 3d $\cN=4$ theory with $SU(2)$ R-symmetry. 
    
\item For the $\cN=(1,0)$ theory, the R-symmetry group is $SU(2)_R \supset U(1)_t$.
After twisting, the supercharges transform as
\begin{eqnarray}
\begin{split}
    SO(6)  \times SU(2)_R & \quad \to  &   SU(2)_{\bR^3}  \times U(1)_{\text{tw}}, \\
    \bf{(4^+,2)}  & \quad \to  & 
    \bf{2_0}  +  \bf{2_0}+  \bf{2}_{\pm 2}.
\end{split}
\end{eqnarray}
There are four supercharges leading to a 3d $\cN=2$ theory. 
\end{itemize}

In the rest of this section, we will study the compactifications of 6d $\mathcal N=(1,0)$ theories on non-compact 4-manifolds, which bring forth various 3d-2d coupled systems, and their gluing to compact ones. In the case of MSW twist on K\"ahler 4-manifolds with boundaries, the 2d theories turn out to admit $\mathcal N=(0,2)$ supersymmetry. In comparison with the usual setup of 3d $\mathcal N=2$ theories with $(0, 2)$ boundaries, there are some curious observations here based on our previous analysis that the 3d TQFTs, which are reached via RG flow from 3d $\mathcal N=1$ theories upon compactification, are bounded by 2d $\mathcal N=(0, 2)$ theories with a half-twist on the right-moving sector. It would be interesting to have a better understanding of this phenomenon. However, we will not pursue this goal in the current work.

\subsection{Gluing at the level of geometry} 
\label{subsec:4.2}

\begin{figure}
    \centering
    \includegraphics{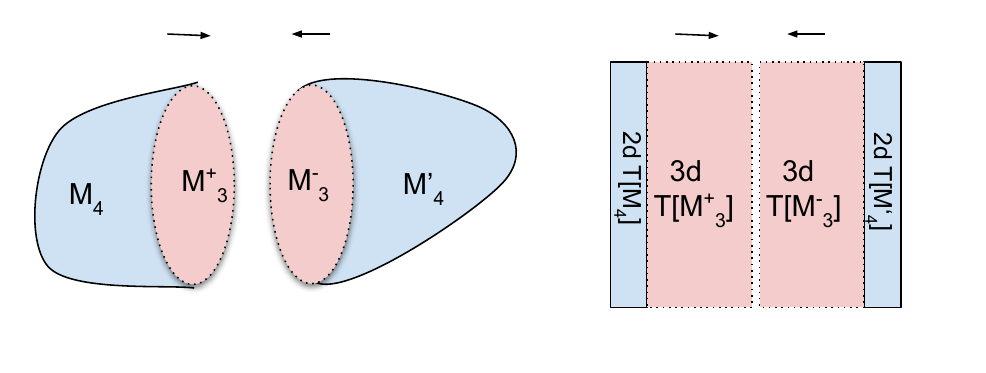}
    \caption{Two non-compact 4-manifolds are glued along their common boundary to a compact 3-manifold. At the field theory level, the coupled 3d-2d system are fused to a pure two-dimensional SCFT.}
    \label{fig:compactgluing}
\end{figure}

The compactification on non-compact 4-manifolds in general leads to a coupled 3d-2d system. 
Although we can study the 2d theory $T[M_4]$ and 3d theory $T[M_3]$ individually, how to couple them together into a consistent system is complicated. 
The known examples include 6d abelian theories and a few others. 
In this section, we will study the gluing of the non-compact 4-manifolds along their common boundaries. 
Two non-compact 4-manifolds can be glued together in such a way that a new coupled 3d-2d system arises which defines a fusion at the level of the 2d SCFTs. Similarly, two non-compact 4-manifolds with the same 3-manifold boundary $M_3$ of opposite orientation can be glued to a compact manifold, and the coupled 2d-3d systems fuse together to a pure 2d SCFT. This procedure is shown schematically in Figure \ref{fig:compactgluing}.
We will study how this gluing of theories works at least at the level of the chiral algebra using anomaly polynomials.

The general principle is that the total anomaly polynomial of the theories before and after the gluing should be the same. 
The anomaly polynomial or central charges for the non-compact spaces are usually computed equivariantly with parameters $\epsilon_{1,2}$ in their expression, while for compact spaces, the computation of the anomaly polynomials are straightforward and the results are independent of these parameters
\footnote{Note that introducing equivariant parameters regularize the geometry and the boundaries do not directly contribute to dimensional reduction of the anomaly polynomial. 
The central charge and anomaly polynomials are still given by the ones for compact 4-manifolds with $\chi$ and $\sigma$ replaced by their equivariant version.
}. 
Thus, if we are doing the gluing properly, the glued anomaly polynomial should be independent of $\epsilon_{1,2}$ and equal to the anomaly polynomial of the corresponding compact one. 
We will see more of these in examples below.

The gluing rule for toric 4-manifolds $M_4$ has been studied in \cite{Fucito:2006kn,Griguolo:2006kp,Dijkgraaf:2007fe, Cirafici:2009ga,Gasparim:2009sns,Bonelli:2011jx, Bonelli:2011kv,Ito:2011mw,Alfimov:2011ju,Bonelli:2012ny,Bruzzo:2009uc,Bruzzo:2013daa,Bruzzo:2014jza,Bawane:2014uka,Bershtein:2015xfa, Bershtein:2016mxz, Feigin:2018bkf}. 
The idea is that toric 4-manifolds have a $U(1)^2$ torus action, which descends to the $U(1)^2$ action on local $\bC^2$ patches.
If we treat $M_4$ as a gluing of its local patches, then from the toric data, one can identify the relations between equivariant parameters $\epsilon_{1,2}$ on each patch such that they glue to $M_4$. We summarize this procedure in Appendex \ref{app:B1}.
With this gluing rule in hand, one can for example compute the instanton partition functions $Z$ of 4D gauge theories  on both non-compact \cite{Fucito:2006kn,Griguolo:2006kp,Dijkgraaf:2007fe, Cirafici:2009ga,Gasparim:2009sns,Bonelli:2011jx,  Bonelli:2011kv,Ito:2011mw,Alfimov:2011ju,Bonelli:2012ny,Bruzzo:2009uc,Bruzzo:2013daa,Bruzzo:2014jza,Bawane:2014uka,Bershtein:2015xfa, Bershtein:2016mxz} and compact space
\cite{Bawane:2014uka,Bershtein:2015xfa, Bershtein:2016mxz} by first evaluating $Z$ on each patch $\bC^2$ and then glue the results together using the gluing rule for $M_4$.

In the spirit of the AGT correspondence, one can also study the gluing of the chiral algebra via the central charges and anomaly polynomials \cite{Bonelli:2012ny, Hosseini:2020vgl, Feigin:2018bkf}.
For the toric 4-manifolds, the basic building block is $\bC^2$, the chiral algebra in the 2d SCFT is in general the direct sum of $W$ algebras. 
Besides that, one can study the gluing of two 4-manifolds if and only if they share the same boundary. 
We will not restrict to toric 4-manifolds, but study the general gluing rule for a large class of 4-manifolds constructed from plumbing. We expect more diverse realizations of the chiral algebra apart from sums of W-algebras.

\subsubsection*{Gluing of toric 4-manifolds from local patches}

\paragraph{Example: $\bR^4$} \mbox{}\\
The first non-compact 4-manifold that we will consider is $\bR^4$. Equivariantly, it is treated as a 4-ball $B^4_{\epsilon_{1,2}}$ where $\epsilon_1$ and $\epsilon_2$ are equivariant parameters associated with the isometry $U(1)^2$. 
As a toric manifold, it can be represented by two complex lines $\bR^2_{\epsilon_i} \cong \bC_{\epsilon_i}$ fixed by the $U(1)$ factors. The boundary is just $\partial B^4=S^3 $.

The compactification of the 6d theories on the non-compact 4-manifold $\bR_{\epsilon_{1,2}}$ leads to a 3d-2d coupled system with $T[S^3]$ in the bulk and $T[\bR^4_{\epsilon_{1,2}}]$ on the boundary. Most of the time, it is difficult to determine the theory $T[S^3]$. But, with the help of anomaly polynomial reductions, we know the central charge of $T[\bR^4_{\epsilon_{1,2}}]$ and thus the gravitational anomaly of $T[S^3]$.

The equivariant Euler number and signature can be calculated using the localization formula (\ref{eqn:geo_formula}). For $\bR^4$, there is only one fixed point. Thus, the results are 
\begin{equation} \label{eqn:geo_r4}
    \tilde{\chi}(\bR^4) = 1, \quad \tilde{\sigma}(\bR^4) = \frac{1}{3} \frac{\epsilon_1^2+\epsilon_2^2}{\epsilon_1 \epsilon_2} = \frac{1}{3} (\alpha + \frac{1}{\alpha}) =  \frac{1}{3} ((b+\frac{1}{b})^2-2).
\end{equation}
Here, we introduce the parameter $\alpha = b^2=\epsilon_2/\epsilon_1$ to encode the equivariant parameters. This will be one of the building blocks to construct more general 4-manifolds by gluing.

\paragraph{Example: $\bP^2$}  \mbox{}\\ 
Let's consider $\bP^2$ as an example of a compact toric 4-manifold. The toric data is given in terms of vertices of the toric fan,
\begin{equation*}
    v_0=(1,0), \quad v_1=(0,1), \quad v_2=(-1,-1).
\end{equation*}
Using the equation (\ref{eqn:v2alpha}), one finds the relation of equivariant parameters between different patches
\begin{equation} \label{eqn:alpha_p2}
    \alpha_1 = \frac{\alpha}{\alpha-1}, \qquad \alpha_2 = 1- \alpha, \quad \alpha_3=\alpha.
\end{equation} 
Notice that these parameters satisfy the monodromy free condition $\alpha_1+\alpha_2^{-1} = 1$. 
Plugging this into the equivariant geometric data of $\bR^4$ in \eqref{eqn:geo_r4}, 
we find that 
\begin{equation}
    \chi(\bP^2) = 3, \quad \sigma(\bP^2) = \frac{1}{3}(\alpha_1+\frac{1}{\alpha_1}+\alpha_2+\frac{1}{\alpha_2}+\alpha_3+\frac{1}{\alpha_3}) =1,
\end{equation}
which agree with the Euler number and signature of $\bP^2$.

\paragraph{Example: $\cO_{\bP^1}(-p)$} \mbox{}\\
As an example of a non-compact toric 4-manifold, we consider the line bundle $\cO_{\bP^1}(-p)$ which is the resolution of the singular $\bC^2/\bZ_p$ surface while $\bZ_p$ acting as 
\begin{equation*}
(z_1,z_2) \to \omega (z_1,z_2), \quad \omega = \exp{(2\pi i /p)}. 
\end{equation*}
By Hirzebruch-Jung resolution discussed in appendix \ref{app:B1}, one can show that 
\begin{equation} \label{eqn:alpha_op}
   \alpha_1 = \frac{p\alpha}{1-\alpha},  \qquad \alpha_2 = \frac{\alpha-1}{p}.
\end{equation}
Thus, the equivariant Euler number and signature are given by
\begin{equation} \label{eqn:geo_op}
    \tilde{\chi}(\cO_{\bP^1}(-p)) = 2, \quad \tilde{\sigma}(\cO_{\bP^1}(-p)) =\frac{1}{3p}\left( \alpha + \frac{1}{\alpha} -(p^2+2)\right).
\end{equation}

\subsubsection*{Gluing along a common boundary}

We have seen how the toric 4-manifolds are glued together from local patches utilizing the toric datas. Next, we'd like to show how different non-compact toric 4-manifolds can be further glued together along their common boundary. We will consider 4-manifolds $M_4$ constructed by plumbing disk bundles \cite{Gadde:2013sca}. 
For simply connected 4-manifolds without 1-cycles, they can be expressed in terms of plumbing graphs. 
As reviewed in the appendix \ref{app:B2}, the boundary $M_3$ of plumbing 4-manifolds can be calculated from the plumbing graph.

The simplest plumbing 4-manifold is $M_4=\cO_{\bP^1}(-p)$.
It is just one disk bundle with Euler number $p$. The plumbing graph is  $\Upsilon = \left({-p \atop \bullet}\right)$. 
Using the method in appendix \ref{app:B2}, one can find that its boundary is the lens space $L(p,1)$. Recall that the lens spaces $L(p,q)$ are quotients of $S^3 \subset \bC^2$ by a free acting $\bZ_p$ determined by two coprime integers $p$ and $q$ as
\begin{equation*}
(z_{1},z_{2}) \to (e^{2\pi i/p} z_{1},e^{2\pi iq/p}z_{2}). 
\end{equation*}
To glue $\cO_{\bP^1}(-p)$, one needs to find some other 4-manifold also bounded by $L(p,1)$.
We will study the different gluings of $\cO_{\bP^1}(-p)$ in the following.

For non-compact toric 4-manifolds $M_4^+$ and $M_4^-$, their Euler characteristic and signature depend on the equivariant parameters $\alpha_+$ and $\alpha_-$.
If we demand that the 4-manifold after gluing, $M_4=M_4^+ \cup M_4^-$, does not have non-trivial monodromy, then these 
parameters should satisfy
$\alpha_+ + \alpha_-^{-1} =a$ with $a\in \bZ$ \cite{Feigin:2018bkf}. For plumbing manifolds, this integer is the Euler number of the disk bundles used in the construction. 
For example, $\cO_{\bP^1}(-p)$ can be understood as the gluing of two $\bR^4$ with equivariant parameters given in (\ref{eqn:geo_op}). As one can easily check, $\alpha_1 + \alpha_2^{-1} =-p$. 
For more general non-compact plumbing manifolds, we refer to the appendix \ref{app:B2}.

For the case of a simple gluing of two 4-manifolds along their common boundary, since there are no twists involved in this process, the equivariant parameters should satisfy $\alpha_+ + \alpha_-^{-1} =0$. 
Besides this condition, one needs to make sure that $M_4^+$ and $M_4^-$ have opposite orientations on their boundaries. 
Given a 4-manifold $M_4$, we can reverse its orientation simply by switching the roles of $b^2_+ \leftrightarrow b^2_-$ of the lattice \cite{Gadde:2013sca}. We denote the reversed manifold as $\overline{M_4}$. Due to this switch, the signature should be modified as $\sigma(M_4)=-\sigma(\overline{M_4})$. This condition can be also realized for the equivariant signature for non-compact spaces
\footnote{For a local patch $\bR^4_{\epsilon_{1,2}}$, the equivariant signature is $\tilde \sigma(\bR^4)=(\alpha+\alpha^{-1})/3$. Reversing the orientation of $\bR^4$ just amounts to changing the equivariant parameters from $\alpha$ to $-\alpha$ such that $\tilde \sigma(\bR^4_{\epsilon_{1,2}})=-\tilde \sigma(\overline{\bR^4})$.}.

\paragraph{Example: $\cO_{\bP^1}(-1) \cup \bR^4$} \mbox{}\\
When $p=q=1$, the action is trivial and the lens space reduces to $S^3$. In equivariant sense, this is just the boundary of $\bR^4$.
It implies that we can glue $\cO_{\bP^1}(-1) \cup \bR^4$ along their boundary leading to a compact 4-manifold. Taking $p=1$ in equation (\ref{eqn:geo_op}), we get %
\begin{equation*} 
    \tilde{\chi}(\cO(-1)) = 2, \quad \tilde{\sigma}(\cO(-1)) =\frac{1}{3}\left( \alpha + \frac{1}{\alpha}\right)-1 \;.
\end{equation*}
Now, adding them to $\tilde{\chi}(\bR^4)$ and $\tilde{\sigma}(\bR^4) $ in (\ref{eqn:geo_r4}), we get exactly the Euler characteristic and signature of $\overline{\bP^2}$. Thus, the central charge becomes $ c_L(\cO(1)) + c_L(\bR^4)=c_L(\overline{\bP^2})$. 

Similarly, for $p=-1$, we can glue $\cO(1)$ with a $\overline{\bR^4}$ along the common boundary to obtain $\bP^2$. 
In fact, $\cO(1)$ is just $\bP^2/\{pt\}$, i.e. $\bP^2$ with one puncture \cite{Feigin:2018bkf}, and the gluing with $\overline{\bR^4}$ is exactly the operation of closing puncture.

\paragraph{Example: $\cO_{\bP^1}(-p)$ $\cup$ $\overline{\cO_{\bP^1}(-p)}$}  \mbox{}\\
As discussed above, by the relation between the Euler characteristic and signature $ \tilde{\chi}(\cO_{\bP^1}(-p))$ $=  \tilde{\chi}(\overline{\cO_{\bP^1}(-p)})$ and $ \tilde{\sigma}(\cO_{\bP^1}(-p))= - \tilde{\sigma}(\overline{\cO_{\bP^1}(-p)})$, the compact 4-manifold after the gluing, denoted by $M_4$, has the Euler characteristic $\chi(M_4)=4$ and $\sigma(M_4)=0$ with plumbing diagram $\Upsilon = \left(
\raisebox{-0.7ex}{$\overset{-p}{\bullet}\hspace{-0.7em} -\hspace{-0.7em}-\hspace{-0.7em}-\hspace{-0.7em}\overset{p}{\bullet}$}
\right)$. 
In terms of the 2d effective fields, it implies that the there are $b_2=b_2^++b_2^-$ left-/right-moving chiral bosons and can be understood as $b_2$ non-chiral bosons in $T[M_4]$.

\paragraph{Example: $\cO_{\bP^1}(-p)$ $\cup$ $A_{p-1}$}  \mbox{}\\
Besides $\overline{\cO_{\bP^1}(-p)}$, as shown in \cite{Gadde:2013sca}, by Kirby moves, one can show that the boundary of $A_{p-1}$ is  $L(p,-1)$, which is exact the same boundary as the one of $\cO_{\bP^1}(-p)$ with opposite orientation. Thus, we don't need to reverse the orientation when gluing. 

The equivariant Euler characteristic and signature of the $A_{p-1}$-manifold are 
\begin{equation} \label{eqn:geo_ALE}
    \tilde \chi (A_{p-1}) = p, \qquad \tilde \sigma(A_{p-1}) = \frac{1}{3p}(\alpha+\frac{1}{\alpha}+2-2p^2),
\end{equation}
Adding these to $ \tilde \chi (\cO_{\bP^1}(-p)) $ and $ \tilde \sigma (\cO_{\bP^1}(-p)) $ given in equation (\ref{eqn:geo_op}) and taking $\alpha_+ + {\alpha_-}^{-1}=0$, 
we get that 
\begin{equation}
    \tilde \chi ((\overline{\bP^{2}})^{\#^p}) = p+2, \qquad \tilde \sigma((\overline{\bP^{2}})^{\#^p}) = -p,
\end{equation}
which is exactly the same the result as predicted by the Kirby calculus. Here, the connected sum of two compact 4-manifolds means removing a small 4-ball $B^4$ from both manifolds and then gluing them along their common boundary $S^3$.

\subsection{Gluing for 6D $\cN=(1,0)$ SCFTs}

For 6D $\cN=(1,0)$ SCFTs, the anomaly polynomial $I_4$ after the dimensional reduction contains besides the term from the gravitational anomaly and R-current, the terms depending on flavor symmetries. 
For simple conformal matters and class $S_k$ theories, there are two flavor symmetries $G_L$ and $G_R$. 
In the compactification, these flavor symmetries descends to the 2D CFT, which is reflected in the anomaly polynomial $I_4$, as one can see from equation (\ref{eqn:I4_ade}) for conformal matter and equation (\ref{eqn:2dF_sk}) for class $S_k$ by terms propositional to $\Tr{F^2_L}$ and $\Tr{F^2_R}$, where $F_L$ and $F_R$ are the field strengths of the background gauge fields. 
Notice that if there is no flux, the 2D field strength $F$ does not depend on the internal manifold and is the same for any K\"ahler 4-manifold.

Consider the gluing of two non-compact 4-manifolds $M_4^+$ and $M_4^-$ into a manifold $M_4$. 
The anomaly polynomial should be the same before and after the gluing 
\begin{equation*}
    I_4(M_4) =  I_4(M^+_4) + I_4(M^-_4) \;,
\end{equation*}
for 6D $\cN=(1,0)$ SCFTs, which is equivalent to requiring that both the central charges and flavor dependent terms respect the gluing. The correct addition of the central charges should be clear as they only appear through linear terms in the topological invariants of $M_4$ in the anomaly polynomial. We now show that the field strength dependent terms also respect the gluing. To this end, notice that for a 4-manifold $M_4^+$ with 3-manifold boundary $M_3$, the integral of the field strength contributions becomes
\begin{equation} \label{eq:FFinv}
    I_a(M_4^+) \equiv \frac{1}{8\pi}\int_{M_4^+} \mathrm{Tr} F_a^2,
\end{equation}
which for topologically trivial $a$ can be rewritten as\footnote{For topologically non-trivial gauge field $a$, the formula \eqref{eq:FFinv} has to be used.}
\begin{equation}
    I_a(M_4^+) = \frac{1}{8\pi}\int_{M_4^+} d \omega_a = \frac{1}{2\pi}\int_{M_3} \omega_a,
\end{equation}
where $\omega_a$ is the Chern-Simons form,
\begin{equation}
    \omega_a = \mathrm{Tr} \left(\frac{2}{3} a^3 + a \wedge da \right),
\end{equation}
giving the \textit{Chern-Simons invariant} over $M_3$. If now $M_4'^+$ is another 4-manifold with the same boundary $M_3$, then we have
\begin{equation}
    \frac{1}{8\pi}\left(\int_{M_4^+} \mathrm{Tr} F_a^2 - \int_{M_4'^+} \mathrm{Tr} F_a^2 \right) = \frac{1}{8\pi} \left(\int_{M_4^+} \mathrm{Tr} F_a^2 + \int_{M_4^-} \mathrm{Tr} F_a^2\right) = \frac{1}{8\pi} \int_{M_4} \mathrm{Tr} F_a^2~.
\end{equation}
Now since the cohomology calss $[F_a/2\pi]$ is integral, we get 
\begin{equation}
    \frac{1}{8\pi} \int_{M_4} \mathrm{Tr}F_a^2 \in 2\pi \cdot \mathbb{Z}, 
\end{equation}
which simultaneously shows that the fluxes are integrally quantized and that, for given $M_3$, $I_a(M_4^+)$ does not depend on the choice of $M_4^+$ modulo $2\pi \mathbb{Z}$.

For example, consider the worldvolume theory of a single M5 brane probing $\bZ_k$ singularities. From the anomaly polynomial $I_4$ in (\ref{eqn:I4_ade}), and Table \ref{tab:6d2dcc}, one has that 
\begin{equation} \label{eqn:I4_a}
    I_4(M_4) = -\frac{\chi+5\sigma}{96}P_1(T\Sigma) + \frac{\chi+\sigma}{8}C^2_1(R) + 
    \frac{k}{32} \sigma (\tr{F_L^2}+ \tr{F_R^2}),
\end{equation}
where $F_L$ and $F_R$ are field strengths of background gauge fields $SU(k)^2$. 
The left-moving central charge is 
\begin{equation}
    c_L(M_4) = \frac{1}{2} \chi+\frac{k^2-4}{8}\sigma.
\end{equation}
Note that $I_4(M_4)$ depends linearly on the Euler characteristic $\chi$ and signature $\sigma$ and we have 
\begin{equation}
    \chi(M_4) = \tilde \chi(M_4^+) + \tilde \chi(M_4^-),  \quad \sigma(M_4) = \tilde \sigma(M_4^+) + \tilde \sigma(M_4^-) \;, 
\end{equation}
in the gluing of $M_4 = M_4^+ \cup M_4^-$. Thus, the full anomaly polynomial $I_4$ should respect the gluing. 
We will check this using the 6D $\cN=(1,0)$ theory of a single M5 brane probing $\bZ_k$ singularities for several different gluing examples in subsection \ref{subsec:4.2}.

\subsubsection*{Example: $\bR^4$}

Consider the simplest non-compact 4-manifolds $\bR^4$. The equivariant Euler number and signature are given in equation (\ref{eqn:geo_r4}). 
The anomaly polynomial is 
\begin{equation} \label{eqn:2dcc_r4_10}
    \tilde I_4(\bR^4_{\alpha}) = \frac{12C^2_1(R)-P_1(T\Sigma)}{96} + 
    \left(\alpha+\frac{1}{\alpha} \right)\frac{12C^2_1(R)+3k(\tr{F_L^2}+ \tr{F_R^2})-5P_1(T\Sigma)}{288}
     ,
\end{equation}
where $\alpha=\epsilon_2/\epsilon_1$ is the equivariant parameters. As before, $\tilde I_4(\bR^4_{\alpha})$ is to emphasis that the Euler characteristic and signature used in the expression is the equivariant ones. 
The left-moving central charge from $I_4$ is 
\begin{equation}
    c_L=\frac{1}{2} + \frac{k^2-4}{24} (\alpha+\frac{1}{\alpha}) =  \frac{10-k^2}{12}+\frac{k^2-4}{24} (b+\frac{1}{b})^2.
\end{equation}
It is not clear which chiral algebra it is. In the Sec. \ref{sec:MTC}, we will see that the central charge has the same large $k$ behavior with the $k$-th para-Toda theory of type $SU(k)$.

\subsubsection*{Example: $\bP^2$}

Let's consider an example of compact toric 4-manifold $\bP^2$. Since $\chi(\bP^2)=3$ and $\sigma(\bP^2)=1$, using the equation (\ref{eqn:I4_a}), we can compute the anomaly polynomial 
\begin{equation}
    I_4(\bP^2) =  -\frac{1}{12}P_1(T\Sigma) + \frac{1}{2}C^2_1(R) + 
    \frac{k}{32} (\tr{F_L^2}+ \tr{F_R^2}).
\end{equation}
As we discussed in subsection above, $\bP^2$ can be understood as the gluing of three copies of $\bR^4$. 
By direct calculation, one can show that 
\begin{equation*}
I_4(\bP^2) = \tilde I^{(1)}_4(\bR^4_{\alpha_1}) + \tilde I_4^{(2)}(\bR^4_{\alpha_2}) + \tilde I_4^{(3)}(\bR^4_{\alpha_3}),
\end{equation*}
with the equivariant parameters from the equation \eqref{eqn:alpha_p2}. In particular, the left-moving central charge on $\bP^2$ is 
\begin{equation}
    c_L(\bP^2) = \frac{3}{2} +\frac{k^2-4}{8} ,
\end{equation}
and clearly it also respect the gluing of the geometry.

\subsubsection*{Example: $\cO_{\bP^1}(-p)$}

Let's consider an example of non-compact toric 4-manifold $\cO_{\bP^1}(-p)$. 
Plug the equivariant Euler characteristic and signature of $\cO_{\bP^1}(-p)$ from (\ref{eqn:geo_op}) into the equation (\ref{eqn:I4_a}). We have the 2d anomaly polynomial 
\begin{align} \label{eqn:I4_op}
\begin{split}
     I_4(\cO_{\bP^1}(-p))_{\alpha} =&  
    \left( \alpha+\frac{1}{\alpha}-2-p^2 \right)\frac{12C^2_1(R)+3k(\tr{F_L^2}+ \tr{F_R^2})-5P_1(T\Sigma)}{288p}\\
    & \qquad + \frac{12C^2_1(R)-P_1(T\Sigma)}{48}.
\end{split}
\end{align}
Since $\cO_{\bP^1}(-p)$ can be obtained from two patches of the $\bR^4$, we can show that the same anomaly polynomial can be derived by summing up two copies of the $\tilde I_4(\bR^4_{\alpha})$ with
\begin{equation*}
    I_4(\cO_{\bP^1}(-p)) = \tilde I^{(1)}_4(\bR^4_{\alpha_1}) + \tilde I_4^{(2)}(\bR^4_{\alpha_2}),
\end{equation*}
where $\alpha_1$ and $\alpha_2$ are the equivariant parameters on the corresponding patches (\ref{eqn:alpha_op}). Then, the left-moving central charge is 
\begin{align} \label{eqn:cc_op}
\begin{split}
    c_L(\cO_{\bP^1}(-p)) &= 1 +\frac{k^2-4}{24p}(\alpha+\frac{1}{\alpha}-2-p^2) \\
    &= 1-\frac{(p^2+4)(k^2-4)}{24p}+
    \frac{k^2-4}{24p}\left(b+\frac{1}{b}\right)^2 .   
\end{split}
\end{align}

\subsubsection*{Example: $\cO_{\bP^1}(-p)$ $\cup$ $A_{p-1}$}

As an example of the gluing two 4-manifolds along the common boundary, we would like to study the case $\cO_{\bP^1}(-p)\cup A_{p-1}=(\overline{\bP^{2}})^{\#^p}$ which have already shown that the gluing works at the level of geometry in the last section. We will check that the gluing also works at the level of anomaly polynomials. With equivariant geometry data of $A_{p-1}$ in (\ref{eqn:geo_ALE}), the anomaly polynomial is given by
\begin{align*}
\begin{split}
        & I_4(A_{(p-1),\alpha_1}) \\
    &\quad =\left(\alpha_1+\frac{1}{\alpha_1}+2-2p^2 \right) \frac{12C^2_1(R)+3k(\tr{F_L^2}+ \tr{F_R^2})-5P_1(T\Sigma)}{288p}
       + \frac{12C^2_1(R)-P_1(T\Sigma)}{96}p ,
\end{split}
\end{align*}
Add it with $I_4(\cO_{\bP^1}(-p)_{\alpha_2})$ in equation (\ref{eqn:I4_op}) and take into account the monodromy free condition $\alpha_1+\alpha_2^{-1}=0$. The final result is 
\begin{align}
\begin{split}
        &I_4(A_{(p-1),\alpha_1})+ I_4(\cO_{\bP^1}(-p)_{\alpha_2})    \\ 
    &\quad =\frac{5P_1(T\Sigma)-12C^2_1(R)-3k(\tr{F_L^2}+ \tr{F_R^2})}{96}p 
         +\frac{12C^2_1(R)-P_1(T\Sigma)}{96}\left(p+2\right),
\end{split}
\end{align}
which is exactly the anomaly polynomial for $(\overline{\bP^{2}})^{\#^p}$.

The left-moving central charge of $A_{p-1}$ space is 
\begin{align}
\begin{split}
    c_L(A_{(p-1),\alpha_1}) &= \frac{p}{2} +\frac{k^2-4}{24p}(\alpha_1+\frac{1}{\alpha_1}+2-2p^2) \\
    &=  \frac{p(10-k^2)}{12}+
    \frac{k^2-4}{24p}\left(b+\frac{1}{b}\right)^2.
\end{split}
\end{align}
Adding it with the left-moving central charge of $\cO_{\bP^1}(-p)$ in (\ref{eqn:cc_op}), we get
\begin{equation}
    c_L((\overline{\bP^{2}})^{\#^p}) = \frac{p+2}{2} -\frac{k^2-4}{8p} ,
\end{equation}
which is the correct left-moving central charge for $(\overline{\bP^{2}})^{\#^p}$.

Although we have only checked that the anomaly polynomial respect the gluing of geometry using the simplest $\cN=(1,0)$ theories, since the linear dependent with $\chi$ and $\sigma$ in the expression, this gluing formalism of anomaly polynomials should work for other $\cN=(1,0)$ theories.

\section{Concrete CFT proposals}
\label{sec:MTC}

Based on the results from previous sections, we would like to explore how specific chiral algebras arise from compactifications of 6d $(2,0)$ theories and 6d $(1,0)$ theories. Following \cite{Vafa:2015euh}, we will identify the resulting conformal field theories with series of minimal models and Toda theories.

\subsection{$A_{N-1}$ theory on K\"ahler surface}

Let's start by reviewing the case of compactification of the 6d $(2,0)$ $A_1$-type theory on $\bR^4$ with equivariant parameters $\epsilon_{1,2}$ \cite{Alfimov:2013cqa, Vafa:2015euh}. 
According to the AGT correspondence, the corresponding 2d CFT is the Liouville theory with the following action,
\begin{equation*}
    S=\int d^2z\big[ \ {1\over 8\pi}\partial \phi {\overline \partial} \phi +Q R \phi +\mu\  {\rm exp}(2b\phi ) \big].
\end{equation*}
The central charge of this theory is giving by
\begin{equation*}
    c(A_1)=1+6(b+\frac{1}{b})^2 \;.
\end{equation*}

In Liouville theory, there is a special set of fields called degenerate fields given by operators $\Phi_{r,s}$ with momentum
\begin{equation*}
    \alpha = (r-1) b + (s-1)\frac{1}{b}, 
\end{equation*}
for $1\leq r< n, 1\leq s < m$. So, there are totally $mn$ degenerate fields, which will be the anyons in the corresponding 3d bulk theory. 
The OPE of these degenerate fields realize the operator algebra for the minimal model, i.e. 
\begin{equation*}
    \Phi_{r_1,s_1} \times \Phi_{r_2,s_2}=\sum_{\substack{k=1+|r_1-r_2|, k+r_1+r_2+1=0\ {\rm mod}\ 2\\ l=1+|s_1-s_2|, l+s_1+s_2+1=0\ {\rm mod}\ 2}}^{\substack{k=r_1+r_2-1\\ l=s_1+s_2-1}}\Phi_{k,l}.
\end{equation*}
As a non-compact CFT, there are infinitely many operators in the theory. However, when we set the parameter $b$ to specific values, the theory will truncate into certain rational CFTs. As shown in \cite{Vafa:2015euh}, when taking
\begin{equation} \label{eqn:b2_mn}
    b^2 = -\frac{m}{n},
\end{equation}
with $m,n$ being coprime positive integers ensures that the resulting theory is a minimal model. The central charge now becomes 
\begin{equation}
    c=1 - 6\frac{(n-m)^2}{mn} ,
\end{equation}
which is identified with the central charge of the 2d minimal model $(n,m)$.
The corresponding 3d TQFT can then be specified by extracting braiding matrix, as well as $S$ and $T$ matrices from the $(n,m)$ minimal model, resulting in a complete description in terms of an MTC.

Besides the 6d $(2,0)$-theory of $A_1$-type, we can also consider more general models such as the compactification of general 6d $(2,0)$ theories of type $G=A,D,E$. From the central charge, we expect the effective IR theory to be related to Toda theory with $W_G$ algebra. 
The action of Toda theory is 
\begin{equation*}
    \int_{\Sigma} d^2z\big[ {1\over 8\pi}\partial {\vec \phi}\cdot {\overline \partial}{\vec \phi}+i Q{\vec \rho}\cdot {\vec \phi}R+\sum_{j=1}^{r_{G}} {\rm exp}(b\ {\vec e}_j\cdot {\vec \phi})\big] ,
\end{equation*}
where $\vec \phi$ is an $r_G$-dimensional vector parameterizing the Cartan of $G$, $e_j$ are the simple roots, $Q=b+{1\over b}$, $\vec \rho$ is half the sum of positive roots of $G$. 
The central charge is given by 
\begin{equation*}
    c(G)=r_G+12 {\vec \rho}\cdot {\vec \rho}(b+{1\over b})^2 .
\end{equation*}

When compactifying the 6d $A_{N-1}$ on deformed $\bR^4$, the left-moving part of the effective IR theory is expected to be the $A_{N-1}$ Toda theory with the following central charge,
\begin{equation*}
    c(A_{N-1}) = (N-1) +N(N^2-1)(b+\frac{1}{b})^2,
\end{equation*}
where $Q=b+\frac{1}{b}$ and $b^2 = \frac{\epsilon_2}{\epsilon_1}$. Taking $b^2$ to be the same value of (\ref{eqn:b2_mn}), the central charge becomes 
\begin{equation} \label{eqn:cc_todaA}
    c_L = (N-1) - N(N^2-1) \frac{(m-n)^2}{mn},
\end{equation}
which is the central charge for the minimal model $W_N(m,n)$ \cite{Chang:2011vka}. Similar to the Virasoro minimal models, the $W_N(m,n)$ minimal models are parameterised by two coprime integers $m,n>N$ and are unitary if and only if $|m-n|=1$. As in the minimal model case, the corresponding MTC data are determined by the $W_N(m,n)$ models.

\subsubsection*{$A_{N-1}$ theory on $\cO_{\bP^1}(-p)$}

From the results of central charges (\ref{eqn:c20}), we obtain two copies of Liouville theories with the parameters given in the equation (\ref{eqn:alpha_op}). Now take the parameters to be negative rational numbers as follows,
\begin{equation} \label{eq:para}
    b^2 = -\frac{m}{n} \quad 
    b^2_0 = -\frac{m+n}{np}, \quad  
    b^2_1 = -\frac{pm}{n + m},  \quad  
\end{equation} 
where $m,n$ are coprime positive integers. 
The central charges become, 
\begin{align}
\begin{split}
   c_L
    &= 
    \left[1 + 6(b_0+\frac{1}{b_0})^2  \right] +
    \left[1 + 6(b_1+\frac{1}{b_1})^2 \right] \\
    &=
    \left[1 - 6\frac{(n (p-1)-m)^2}{n p (m+n)}  \right] +
    \left[1 - 6\frac{(n-m (p-1))^2}{m p (m+n)} \right]
    .
\end{split}
\end{align}
This coincides with  the central charge of the direct sum of minimal model $(np,m+n)$ and $(mp, m+n)$. As in the case of compactification on $\bR^4$, the anyons are realized as the degenerate fields for each copy of Liouville theory, and the corresponding MTC data should be the same as the direct sum of minimal models $(np,m+n)$ and $(mp, m+n)$.  

It is easy to generaize the result to the compactification of the $A_{N-1}$ theory. We take the same parameters as in (\ref{eq:para}), the central charges become
\begin{align*}
    \begin{split} 
    c_L
    &= 
    \left[(N-1) + N(N^2-1)(b_0+\frac{1}{b_0})^2  \right] +
    \left[(N-1) +  N(N^2-1)(b_1+\frac{1}{b_1})^2 \right] \\
    &=
    \left[(N-1) - N(N^2-1)\frac{(n (p-1)-m)^2}{n p (m+n)}  \right] +
    \left[(N-1) - N(N^2-1)\frac{(n-m (p-1))^2}{m p (m+n)} \right].   
    \end{split}
\end{align*}
The central charge is the same as the central charge of the direct sum of $W_N(np, m+n)$ and $W_N(mp, m+n)$ minimal models.

\subsubsection*{$A_{N-1}$ theory on $A_{p-1}$ ALE space}

Similarly, one can consider the compactification of the $\cN=(2,0)$ theory of $A_{1}$ type on a $A_{p-1}$ ALE space. 
Taking $\epsilon_1$ and $\epsilon_2$ to be coprime numbers $m$ and $-n$ in order to obtain minimal models, now the parameters are 
\begin{equation*}
    b^2 = -\frac{m}{n}, \quad 
    b^2_0 = -\frac{m+(p-1)n}{pn}, \quad  
    b^2_1 = -\frac{2m+(p-2)n}{(p-1)n + m}, \cdots,
    b^2_{p-1} = -\frac{pm}{n+(p-1)m}.
\end{equation*}
With the parameters as above, the central charges can be rewritten as 
\begin{align}
\begin{split}
    c_L
    &= 
    \left[1 + 6(b_0+\frac{1}{b_0})^2  \right] +
    \left[1 + 6(b_1+\frac{1}{b_1})^2 \right]  + \ldots +
    \left[1 + 6(b_{p-1}+\frac{1}{b_{p-1}})^2 \right] \\
    &=
    \left[1 - 6\frac{(m-n)^2}{np(m-n+np))}  \right] +
    \left[1 - 6\frac{(m-n)^2}{(m-n+np)(2m-2n+np)} \right] + \\
    & \qquad \qquad \qquad \qquad \qquad \qquad \qquad \qquad \qquad \qquad  \cdots +
     \left[1 - 6\frac{(m-n)^2}{m p (n-m+mp)} \right] 
    ,
\end{split}
\end{align}
which is the same as the central charge of the sum of minimal models $(m-n+np,np)$, $(2m-2n+np,m-n+np)$, $\ldots$, $(mp,n-m+mp)$.

Repeating the same procedure for the $A_{N-1}$ theory, the central charge becomes 
\begin{align*}
\begin{split}
    c_L
    =& 
    \left[(N-1) - N(N^2-1)\frac{(m-n)^2}{np(m-n+np))}  \right]  \\
    & 
    +\left[(N-1) - N(N^2-1)\frac{(m-n)^2}{(m-n+np)(2m-2n+np)} \right]   \\
    &+ \ldots + \\
    &+
     \left[(N-1) - N(N^2-1)\frac{(m-n)^2}{m p (n-m+mp)} \right] 
    .
\end{split}
\end{align*}
This central charge is identified as the direct sum of $W_N$ minimal models of types $(np, m-n+np)$, $(m-n+np, 2m-2n+np)$, $\ldots$, $(mp,n-m+mp)$.

\subsubsection*{$A_{N-1}$ theory on $\overline{\bP^2}$}

From the discussion in previous section, we know that $\overline{\bP^2}$ can be understood as the gluing of three copies of $\bR^4_{\alpha_{\ell}}$ with $\ell=1,2,3$. The equivariant parameters $\{\alpha_{\ell}\}$ for $\bP^2$ are worked out in (\ref{eqn:alpha_p2}). For $\overline{\bP^2}$, we will reverse the orientation on each patch by $\{\alpha_{\ell}\} \to \{- \alpha_{\ell}\}$. Take the special values for these equivariant parameters, we have 
\begin{equation*} 
    b^2_0 = \frac{m}{n} \quad 
    b^2_1 = -\frac{m+n}{n}, \quad  
    b^2_2 = -\frac{m}{m+n}\;.
\end{equation*} 
where $m,n$ are coprime positive integers. 
The central charge now becomes, 
\begin{align*}
\begin{split}
   c_L(\overline{\bP^2})
    &= 
    \left[1 + 6(b_0+\frac{1}{b_0})^2  \right] +
    \left[1 + 6(b_1+\frac{1}{b_1})^2 \right]  +
    \left[1 + 6(b_2+\frac{1}{b_2})^2 \right] \\
    &=
    \left[1 + 6\frac{(m+n)^2}{mn}  \right]
    +
    \left[1 - 6\frac{n^2}{m  (m+n)}  \right] +
    \left[1 - 6\frac{m^2}{n  (m+n)} \right] = 21
\end{split}
\end{align*}
which reproduce the left-moving central charges for $A_1$ theory on $\overline{\bP^2}$ using the equation (\ref{eqn:c20}). 
From the relationship between the central charges, it seems that the 2d theory $T_{A_1}[\overline{\bP^2}]$ could be the extension of minimal models $(np,m+n)$ and $(mp, m+n)$ with another rational CFT with central charge $1 + 6\frac{(m+n)^2}{mn}$ \footnote{Notice that this construction of $T_{A_1}[\overline{\bP^2}]$ is independent of parameters $m,n$.}. 
Due to $\overline{\bP^2} = O_{\bP^1}(-1) \cup \bR^4$, these two minimal models can also be obtained by the analysis for $O_{\bP^1}(-1)$ case by simply taking $p=1$.

\subsection{Class $S_k$ on K\"ahler surfaces}
\label{sec:SkK}

The second example is to consider the compactification of class $S_k$ wrapping four-dimensional K\"ahler manifolds \cite{Gukov:2018iiq}.
The corresponding 2d effective theory has $\cN=(0,2)$ supersymmetry since the internal space is K\"ahler.
Although there is no 2d-4d correspondence for the compactification of $\cN=(1,0)$ theories, it is possible that there is a similar correspondence, after all the structure of the 2d effective theory has $\cN=(0,2)$ supersymmetry. 
Indeed, as shown in \cite{Mitev:2017jqj}, the spectral curves of the 4d $SU(N)$ gauge theories of class $S_k$ can be reproduced from the 2d CFT weighted current correlation functions of the $W_{Nk}$ algebra.
Here, $W_{Nk}$ stands for the $SU(Nk)$ W-algebra. 
It is also known that the chiral algebra of a $SU(N)$ Toda field theory is $W_{N}$. Therefore, 
it seems that the 2d theory corresponding to $S_k$ class might be a mild modification of Toda field theory by changing the algebra from $W_{N} \to W_{Nk}$. 
We will check this by comparing the central charge.

Consider the 2d CFT obtained from the class $S_k$ theory on $\bR^4$. Plugging the geometric data from (\ref{eqn:geo_r4}) into the equation (\ref{eqn:2dcc_sk}), the central charge is  
\begin{equation} \label{eqn:cc_sk_r4}
    c_L =  \frac{(2-3N) k^2+12N-11}{12} 
    +
    \frac{\left(9 N^3-12 N+4\right)k^2 -1}{24}\left(b+\frac{1}{b}\right)^2 \;.
\end{equation}
Unfortunately, $c_L$ has a complicated dependence on $N$ and $k$. 
For simplicity, we will focus on its asymptotic behavior. 
For large $N$ and $k$, it scales as 
\begin{equation}
c_L \sim \frac{3}{8}N^3k^2 \left(b+\frac{1}{b}\right)^2 \;.
\end{equation}
Clearly, it does not match with the central charge of an $SU(Nk)$ W-algebra. By the equation (\ref{eqn:cc_todaA}), it scales as $c_L\sim N^3K^3$ for large $N$ and $k$. 
Thus, the 2d CFT cannot be a simply $SU(Nk)$ Toda theory.

To match the asymptotic behavior of the central charge $c_L \sim N^3k^2$, 
we conjecture that the 2d CFT obtained from the compactification of class $S_k$ theory is related to the $k$th-para Toda theory with type $SU(Nk)$,
\footnote{It is conjectured in \cite{Nishioka:2011jk} that the $m$-th para-Toda model of type $G$ can be obtained from the compactification of $\cN=(2,0)$ of type $G$ on $\bR^4/\bZ_m$. } coupled to some other coset models. 
The $m$-th para-Toda model of type $G$ is defined as \cite{Nishioka:2011jk}
\begin{equation}
S=S\left(\frac{\hat G_k}{\hat  U(1)^{r_G}}\right) + \int d^2x \left[\partial_\mu \Phi \partial_\mu \Phi +  \sum_{i=1}^{r_G} \Psi_i \bar\Psi_i \exp\left(\frac{b}{\sqrt m}\, \alpha_i\cdot \Phi\right) \right] \ .
\end{equation} 
Here, $\hat G_m/\hat  U(1)^{r_G}$ describes the generalized parafermions $\Psi_i$ of type $G$, $\alpha_i$ are simple roots of $G$, $\Phi$ are $r_G$ free bosons with background charge $(b+1/b)\rho/\sqrt{m}$ with the Weyl vector $\rho$. 
The central charge is given by 
\begin{equation}
c= c\left(\frac{\hat G_m}{\hat  U(1)^{r_G}}\right)+r_G+ \frac{h_G d_G}{m}\left(b+\frac1b\right)^2.\label{eqn:cc_para}
\end{equation}
For $m=1$ this is the usual affine Toda theory. From the equation (\ref{eqn:cc_para}), for $G=SU(Nk)$, the corresponding central charge is 
\begin{equation}
    c = \frac{N^3k^2}{N+1} + (N^3k^2 - N) \left(b+\frac{1}{b}\right)^2  .
\end{equation}
In this model, one can reproduce the correct asymptotic behavior $c \sim N^3k^2$ for large $k$ and $N$.
More work needs to be done to find a 2d CFT that matches the exact $c_L$.

\section{Conclusion and outlook}
\label{sec:conclusions}

In this paper we have examined compactifications of 6d $\mathcal{N}=(1,0)$ SCFTs on K\"ahler manifolds while we have focused on the conformal matter class. We have shown that a suitable twist can be employed which preserves two supercharges of same chirality in the remaining two spacetime dimensions. These theories flow to SCFTs in 2d whose central charges we computed by reducing anomaly 8-forms of the corresponding 6d theories. The results from a single M5 brane probing an ADE singularity are summarized in Table \ref{tab:6d2dcc} and equation \eqref{eqn:2dcc_ade}. One can see the left-moving central charge scales as $\sim k^2$ for theories arising from $A_{k-1}$ and $D_k$ singularities. We explain this behaviour by realizing the corresponding compactifications in M-theory on Calabi-Yau fourfolds. The fourfolds have ADE singularities in their fiber and their base is given by the K\"ahler surface in question. Turning on $G$-flux leads to a setup with M5 branes wrapping the K\"ahler surface giving rise to domain walls in the remaining 3d $\mathcal{N}=2$ supersymmetric theory. Counting vacua on the left and right sides, one finds that the number of domain walls connecting them scales as $k^2$ in accordance with the result from the anomaly polynomial reduction. In the future, it would be desirable to have a concrete CFT description for the 2d theories thus obtained. We make some progress towards this direction in Section \ref{sec:SkK} where we observe that the scaling behaviour of 2d central charges obtained by compactifying 6d class $S_k$ theory on K\"ahler surfaces is identical to the scaling of $k$-th para-Toda theories of type $SU(Nk)$. More investigation needs to be done to pin down the CFT more precisely here and to identify the relevant CFTs for D and E type conformal matter theories. A novelty of the 6d conformal matter compactifications as compared to 6d non-Higgsable clusters (or 6d $(2,0)$ theories) is that the anomaly polynomial depends on flavor symmetry field strenghts which can be given flux along the 4-manifold. This will lead to $U(1)$ symmetries in the effective 2d theory and one would need to employ c-extremization to compute the correct central charge. In this paper we have chosen to set all such fluxes to zero and leave the c-extremization problem for future study.

The second part of the paper dealt with compactifications along non-compact K\"ahler manifolds with 3-manifold boundaries and we employed a regularization scheme to compute Euler number and signature of such manifolds equivariantly. The resulting central charges then depend on the equivariant parameters. We then showed how two non-compact 4-manifolds can be glued together using either gluing along toric fans, or alternatively gluing along common 3-manifold boundaries with opposite orientation. In the second case, the resulting 4-manifold is always compact and we show that the central charges add correctly together to reproduce the central of the compact manifold which is independent of equivariant parameters. An important question is about the effective field theory description after compactification on such non-compact 4-manifolds. We have proposed, in analogy to previous work on 6d $(2,0)$ compactifications, that the resulting theory is a coupled 3d-2d system where the 3d theory is the one obtained from compactification on the boundary $M_3$. We have shown that the corresponding 3d theory has $\mathcal{N}=1$ supersymmetry and have proposed that it flows to a topological field theory in the IR. The details of these 3d theories, however, remain unclear and it would be desirable to obtain Lagrangian descriptions of such theories. A concrete path to such a description is available for Seifert manifolds which admit a circle fiber, where one could first reduce along the circle to obtain a 5d supergravity description along the lines of \cite{Cordova:2013bea,Cordova:2013cea}. 

\subsection*{Acknowledgments}
We would like to thank Shi Cheng, Sergei Gukov, Ken Kikuchi, Jianfeng Lin, Mauricio Romo, and Yinan Wang for valuable discussions. This work is supported by the National Thousand-Young-Talents Program of China.  The work of JC is supported by the Fundamental Research Funds for the Central Universities (No.20720230010) of China. The work of WC is supported by the fellowship
of China Postdoctoral Science Foundation NO.2022M720507 and in part by the Beijing
Postdoctoral Research Foundation. WC also would like to thank the organizers of the 2022 BIMSA Workshop on String Theory and the Tsinghua Sanya International Mathematics Forum (TSIMF) for hospitality where part of this work was completed. 

\appendix

\section{$6$D anomaly polynomials}
\label{sec:AA}

The anomaly 8-froms for all three multiplets are given by \cite{Ohmori:2014kda}
\begin{itemize}
    \item  A hypermultiplet in representation $\rho$:
 \begin{align}
  I_{8}^{\mathrm{hyper}} = \frac{1}{24} \Tr_\rho F^4 + \frac{1}{48} \Tr_\rho 
F^2\
p_1(T) + \frac{d_\rho }{5760} \left(7p_1^2(T) -4p_2(T) \right) ,
 \end{align}
    \item A vector multiplet of gauge group $G$:
    \begin{align}
\begin{aligned}
 I_{8}^{\mathrm{vector}} = 
 &-\frac{1}{24} \left( \Tr_{\mathrm{adj}} F^4 + 6 
c_2(R) \Tr_{\mathrm{adj}} F^2 + d_G c_2(R)^2 \right) \\
&-\frac{1}{48} \left( \Tr_{\mathrm{adj}} F^2  + d_G c_2(R)\right) p_1 (T) 
-\frac{d_G }{5760} \left( 7p_1^2 (T) - 4p_2 (T) \right),
\end{aligned}
\end{align}
\item A tensor multiplet:
\begin{align}
  I_{8}^{\mathrm{tensor}} = \frac{1}{24} c_2^2(R) + \frac{1}{48} c_2(R) 
p_1(T) + \frac{1 }{5760} \left(23 p_1^2(T) -116 p_2(T) \right) .
 \end{align}
 \end{itemize}
Here, $d_\rho$ is the dimension of the representation $\rho$, $d_G$ is the dimension of $G$, and the subscripts $\rho$, $\mathrm{f}$, $\mathrm{adj}$ in the trace indicate that it is performed in the representation of $\rho$, adjoint, or  fundamental.

\section{Reduction of anomaly polynomial for E-string theories}

The E-string theory has flavor symmetry $E_8$ for rank one and $SU(2) \times E_8$ for rank higher than one. We will use the notation $SU(2)_R$ for R-symmetry and $SU(2)_L$ for the flavor symmetry.
The anomaly polynomial of the rank $N$ E-string theory is given by \cite{Ohmori:2014pca}
\begin{align}
    &I_{8} = \\
    & \qquad \frac{N(4N^2 + 6N +3)}{24}C^2_2 (R) + \frac{(N-1)(4N^2 - 2N +1)}{24}C^2_2 (L) 
    - \frac{N(N^2 - 1)}{3} C_2 (R) C_2 (L)   \nonumber \\
    &\qquad +  \frac{(N-1)(6N+1)}{48} C_2 (L) p_1 (T) - \frac{N(6N+5)}{48} C_2 (R) p_1 (T)  
    + \frac{N(N-1)}{120} C_2 (L) C_2(E_8)_{\bf{248}}  \nonumber \\
    &\qquad -  \frac{N(N+1)}{120} C_2 (R) C_2(E_8)_{\bf{248}} 
    + \frac{N}{240} p_1 (T) C_2(E_8)_{\bf{248}} + \frac{N}{7200} C^2_2(E_8)_{\bf{248}}   \nonumber \\
    & \qquad +  (30N-1)\frac{7p_1 (T) - 4p_2 (T)}{5760}, \nonumber
\end{align}
where $p_1 (T), p_2 (T)$ are the first and second Pontryagin classes, $C_2 (R), C_2 (L)$ are the second Chern classes in the fundamental representation of the $SU(2)_R$ and $SU(2)_L$ symmetries, and $C_2(E_8)_{\bf{248}}$ is the second Chern class of the $E_8$ flavor symmetry, evaluated in the adjoint representation.

The dimensional reduction of this anomaly polynomial over a K\"ahler surface is studied in Section \ref{sec:MSWtwist2}. The 2d anomaly polynomial has the form of (\ref{eqn:I4-02}), where the central charges are 
\begin{align}
&c_L= \left(36 N^3+90 N^2+87 N-1\right) \frac{\sigma}{8} + N \left(6 N^2+12N+7\right) \frac{\chi}{2},
\nonumber \\
&c_R = \frac{3N}{4}  \left[\left(6 N^2+12 N+7\right) \sigma + \left(4 N^2+6 N+3\right)
  \chi \right].
\end{align}
and the flavor dependent terms are 
\begin{align}
   I_4(F^2) =&
    \left(
    \frac{N(N+1)}{240}C_2(E_8) + \frac{N(N^2-1)}{6}C_2(L)\right)\chi  \\
    &+\left( 
    \frac{ (N+3)N}{160} C_2(E_8)+
    \frac{(4N^2+10N+1)(N-1)}{16} 
    C_2(L) 
    \right)
  \sigma~. \nonumber
\end{align}

Next, consider 2d CFT obtained from the rank $N$ E-strings theory on $\bR^4$. With the geometric data  (\ref{eqn:geo_r4}), the central charge is   
\begin{equation} \label{eqn:cc_Estring_r4}
    c_L =  
    \frac{1 - 45 N - 18 N^2}{12} 
    +
     \frac{36 N^3+90 N^2+87 N-1}{24}\left(b+\frac{1}{b}\right)^2
     \;.
\end{equation}

\section{Four-manifold with boundary}

In this work, we consider compactifications of the 6d SCFTs over 4-manifolds. 
To be specific, we are interested in 4-manifolds with boundaries where we will have a 3d/2d coupled system after compactification. 
Let's review here the constructions and some basic facts about these 4-manifolds with boundaries following \cite{Gadde:2013sca}.
The basic topological invariants of a (compact) 4-manifold $M_4$ are the Betti numbers $b_i (M_4)$.
The manifolds that we will be using are simply-connected ones, i.e. $b_0 (M_4) = 1$. 
They come with a boundary $M_3=\partial M_4$, so that we have $b_4=0$.
We also require $M_3$ to be closed which implies that $b_3=1$ and we require that $b_1 (M_4) = 0$. 
Thus, for the simply-connected 4-manifold with boundary that we will be interested in, the only non-trivial Betti number of $M_4$ is $b_2 \neq 0$.

On the second homology lattice $\Gamma \; = \; H_2 (M_4; \bZ) / \text{Tors}$, one can define a nondegenerate symmetric bilinear integer-valued form by
\begin{equation}
    Q_{M_4} : \Gamma \otimes \Gamma \; \to \; \bZ,
\label{intmatdef}
\end{equation}
which is called the intersection form $Q$ for $M_4$. 
Obviously, the rank of $Q$ is $b_2$. 
Let $b_2^+$ ($b_2^-$) be the number of positive(negative) eigenvalues of $Q$, i.e. $b_2  =  b_2^+ + b_2^-$. 
The Euler characteristic and the signature of $M_4$ are given by
\begin{equation} \label{eqn:chi/sigma}
     \chi  =  2 + b_2^+ + b_2^- , \quad \sigma =  b_2^+ - b_2^- .
\end{equation}
These are the two topological invariants that will play an important role in determining the central charge of $T[M_4]$.

\subsection{Toric 4-manifolds} 
\label{app:B1}

A toric 4-manifold $M_4$ can be described by a set of vectors $\{ v_\ell \}$ with $\ell=1,2,\ldots, n$ in the lattice $N= \bZ^2$. The vectors $v_\ell$ satisfy the relations
 \begin{equation*}
    v_{\ell-1}+v_{\ell+1} - h_{\ell} \, v_{\ell} = 0, \qquad \ell=1,\ldots ,  n  .
\end{equation*}
Notice that only $n-2$ of these relations are independent. Each vector $v_{\ell}$ is associated with a divisor $D_{\ell} \in H_2(M_4,\bZ)$. The intersection form $Q_{M_4}$ is determined by 
 \begin{equation*}
 D_{\ell} \cdot D_{\ell}=-h_\ell  \,   ,  \qquad  
  D_{\ell} \cdot D_{\ell+1} =   D_{\ell+1} \cdot D_{\ell}=1 \;. 
 \end{equation*}

The adjacent vectors $(v_{\ell},v_{\ell+1})$ generate a cone $\sigma_{\ell}$ in $N_{\bR} = N \otimes \bR$. Each such cone corresponds to a local patch of $M_4$ denoted by $U_{\sigma_l}$. 
Let $N^*$ be the dual lattice of $N$ with natural paring $\langle w,u \rangle \in \bZ$. 
The functions on $U_{\sigma_l}$ are determined by the dual cone
\begin{equation*}
    \sigma^{*}_{\ell} = \{w \in N^*_{\bR}| \langle w,u \rangle \geq 0,  \forall u \in  \sigma_{\ell} \},
\end{equation*}
where $N^*_{\bR}=N^* \otimes \bR$. Let $v^*_{\ell}$ and $v^*_{\ell+1}$ be the generator of the dual cone $\sigma^{*}_{\ell}$. The local coordinates on $U_{\sigma_{\ell}}$ are given by 
is 
\begin{equation*}
    z_1^{\ell}= z_1^{v^*_{l,1}} z_2^{v^*_{l,2}}, \qquad 
    z_2^{\ell}= z_1^{v^*_{l+1,1}} z_2^{v^*_{l+1,2}}.
\end{equation*}

Consider a torus action $(z_1,z_2)\to (e^{i\epsilon_1} z_1, e^{i\epsilon_2} z_2)$, which descends to the action on the patch $U_{\sigma_{\ell}}$ by
\begin{equation*}
    \epsilon_1^{\ell} = v^*_{l} \cdot \epsilon, \qquad \epsilon_2^{\ell} = v^*_{l+1} \cdot \epsilon.
\end{equation*}
For a vector $v_\ell=(v_\ell^1,v_\ell^2)^T$, one can find the dual vector to be $v^*_\ell=(v_\ell^2,-v_\ell^1)^T$. With this relation, the equivariant parameters can be written as 
\begin{equation} \label{eqn:v2alpha}
    \epsilon_1^{\ell} = -\det(v_{\ell+1}, \epsilon), \qquad \epsilon_2^{\ell} = \det(v_{\ell}, \epsilon) \;.
\end{equation}
Thus, given the toric data $v_{\ell}$ of $M_4$, one can derive the equivariant parameters on each patch $U_{\sigma_{\ell}}$. 

For toric 4-manifolds $M_4$, if there are only isolated fixed points under the isometry group $U(1)^2$, then the integral of cohomology classes over $M_4$ can be calculated by the localization formula. 
For example, the Euler characteristic and the signature used extensively in this paper can be calculated by  
\footnote{For the derivation of this expression and more general discussion on the application of localization formula, we refer to \cite{Cordes:1994fc,Pestun:2016zxk,Bah:2019rgq,Hosseini:2020vgl}.}
\begin{equation} \label{eqn:geo_formula}
\tilde \chi(M_4)
     =     \sum_{\ell=0}^{n-1} 1 = n, ~~~~
\tilde \sigma(M_4)
     =     \frac{1}{3}\sum_{\ell=0}^{n-1} \frac{(\epsilon_{1}^{\ell})^{2} 
         + (\epsilon_{2}^{\ell})^{2}}{\epsilon_{1}^{\ell}\epsilon_{2}^{\ell}}
         =\frac{1}{3}\sum_{\ell=0}^{n-1} 
         \left(\alpha_{\ell} + \frac{1}{\alpha_{\ell}}\right),
\end{equation}
where $n$ is the number of the fixed points of the torus action $\bC^2$ and $\alpha_{\ell}=\epsilon^{\ell}_2/\epsilon^{\ell}_1$. Here the tilde is to distinguish that the Euler characteristic and signature are calculated using the equivariant cohomology, which is the same as the usual $\chi(M_4)$ and $\sigma(M_4)$ when the space is compact.

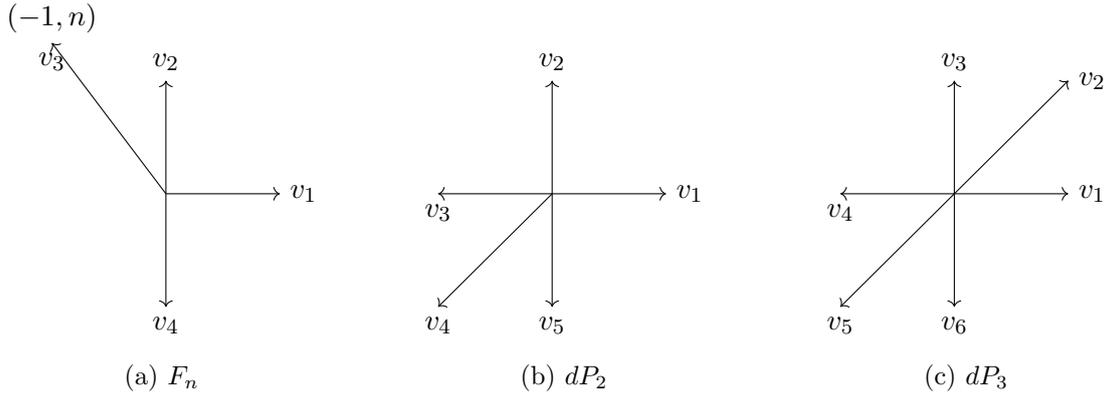
\begin{figure}
     \centering
     \begin{subfigure}[b]{0.3\textwidth}
         \centering
\begin{tikzpicture}
[scale=0.5 ]
\draw[->] (0,0) -- (3,0) node[right] {${ v}_1$};
\draw[->] (0,0) -- (0,3) node[above] {${ v}_2$};
\draw[->] (0,0) -- (-3,4) node[ below] {${ v}_3$} node[above] {$(-1,n)$};
\draw[->] (0,0) -- (0,-3) node[ below] {${ v}_4$};
\end{tikzpicture}
         \caption{$F_n$}
         \label{fig:Fn}
     \end{subfigure}
     \hfill
     \begin{subfigure}[b]{0.3\textwidth}
         \centering
\begin{tikzpicture}
[scale=0.5 ]
\draw[->] (0,0) -- (3,0) node[right] {${ v}_1$};
\draw[->] (0,0) -- (0,3) node[above] {${ v}_2$};
\draw[->] (0,0) -- (-3,0) node[ below] {${ v}_3$};
\draw[->] (0,0) -- (-3,-3) node[ below] {${ v}_4$};
\draw[->] (0,0) -- (0,-3) node[ below] {${ v}_5$};
\end{tikzpicture}
         \caption{$dP_2$}
         \label{fig:dp2}
     \end{subfigure}
     \hfill
     \begin{subfigure}[b]{0.3\textwidth}
         \centering
\begin{tikzpicture}
[scale=0.5 ]
\draw[->] (0,0) -- (3,0) node[right] {${ v}_1$};
\draw[->] (0,0) -- (3,3) node[right] {${ v}_2$};
\draw[->] (0,0) -- (0,3) node[above] {${ v}_3$};
\draw[->] (0,0) -- (-3,0) node[ below] {${ v}_4$};
\draw[->] (0,0) -- (-3,-3) node[ below] {${ v}_5$};
\draw[->] (0,0) -- (0,-3) node[ below] {${ v}_6$};
  \end{tikzpicture}
         \caption{$dP_3$}
         \label{fig:dp3}
     \end{subfigure}
        \caption{Toric diagrams for Hirzebruch surface $F_n$ and Del Pezzo surfaces $dP_2$ and $dP_3$.}
        \label{fig:three graphs}
\end{figure}

\paragraph{Example: Hirzebruch surface}

The toric data of Hirzebruch surface $F_n$ is given in Figure \ref{fig:Fn} with
\begin{equation*}
    v_1=(1,0), \quad v_2=(0,1), \quad v_3=(-1,n), \quad v_4=(0,-1) \;.
\end{equation*}
Using the equation (\ref{eqn:v2alpha}), the equivariant parameters are related by 
\begin{equation*}
    \alpha_1 = \alpha, \quad \alpha_2 = -\frac{1}{n+\alpha}, \quad \alpha_3 = n+\alpha, \quad \alpha_4 = -\alpha~.
\end{equation*}
After equivariant integration using (\ref{eqn:geo_formula}), the Euler characteristic and signature are 
\begin{align*}
    \chi(F_n) =  \sum_{i=1}^4 \tilde{\chi}(\bR_{\alpha_i}^4)=4 \;, 
    \quad \sigma(F_n) = \sum_{i=1}^4 \tilde{\sigma}(\bR^4_{\alpha_i}) = 0~.
\end{align*}

\paragraph{Example: Del Pezzo surfaces}

The Del Pezzo $dP_n$ are the blow-up of $\bC\bP^2$ at $n$ generic points. Note that $dP_0$ is just a $\bP^2$ and $dP_1$ is the Hirzebruch surface $F_1$ studied above. 
We will start from $dP_2$. The toric data of $dP_2$ is given in Figure \ref{fig:dp2} with,
\begin{equation*}
    v_1=(1,0), \quad v_2=(0,1), \quad v_3=(-1,0), \quad v_4=(-1,-1). \quad v_5=(0,-1),
\end{equation*}
Using the equation (\ref{eqn:v2alpha}), the equivariant parameters are related by 
\begin{equation*}
    \alpha_1 = \alpha, \quad \alpha_2 =  -\frac{1}{\alpha}, \quad \alpha_3 = \frac{\alpha}{1-\alpha}, \quad \alpha_4 = \alpha-1, \quad
    \alpha_5 = -\frac{1}{\alpha}~.
\end{equation*}
After equivariant integration using (\ref{eqn:geo_formula}), the Euler characteristic and signature are
\begin{align*}
    \chi(dP_2)  = \sum_{i=1}^5 \tilde{\chi}(\bR_{\alpha_i}^4) = 5 \;,
    \quad 
    \sigma(dP_2) = \sum_{i=1}^5 \tilde{\sigma}(\bR^4_{\alpha_i}) = -1~.
\end{align*}

The next non-trivial example is the $dP_3$. Its toric data is plotted in Figure \ref{fig:dp3} with
\begin{equation*}
    v_1 = (1,0), \quad
    v_2 = (1,1), \quad
    v_3 = (0,1), \quad 
    v_4 = (-1,0), \quad 
    v_5 = (-1,-1), \quad 
    v_6 = (0,-1).
\end{equation*}
Using the equation (\ref{eqn:v2alpha}), the equivariant parameters are related by 
\begin{equation*}
    \alpha_1 = \frac{\alpha}{1-\alpha}, \quad 
    \alpha_2 =  \alpha-1, \quad 
    \alpha_3 =  -\frac{1}{\alpha}, \quad 
    \alpha_4 = \frac{\alpha}{1-\alpha}, \quad 
    \alpha_5 = \alpha-1, \quad
    \alpha_6 = -\frac{1}{\alpha}.
\end{equation*}
After equivariant integration using (\ref{eqn:geo_formula}), the Euler characteristic and signature are
\begin{align*}
    \chi(dP_3)  = \sum_{i=1}^6 \tilde{\chi}(\bR_{\alpha_i}^4) = 6 \;,
    \quad 
    \sigma(dP_3) = \sum_{i=1}^6 \tilde{\sigma}(\bR^4_{\alpha_i}) = -2~.
\end{align*}
To the authors' knowledge, there are no purely toric descriptions for del Pezzo surfaces $dP_n$ with $n>3$.

\subsubsection*{Hirzebruch-Jung resolution}

Consider a class of non-compact 4-manifolds realized as the resolution of the quotient space $\bC^2/\bZ_p$. The action of it depending on two coprime integers $(p,q)$ with $q<p$ is given by 
\begin{equation}
    ( z_1,z_2 ) \to ( e^{2\pi i/p}z_1, e^{2\pi i q/p}z_2 ),
\end{equation}
where $z_1,z_2$ are local coordinates of $\bC^2$. Obviously, this orbifold action has a singularity at the origin of $\bC^2$.

One can resolve the singularities by the Hirzebruch-Jung resolution. The resolved space $X_{p,q}$ contains $n$ exceptional divisors at the origin. The intersection numbers of these divisors are given by 
\begin{equation}
Q \; = \;
\begin{pmatrix}
e_1 & 1   & 0 & \cdots  & 0 \\
1   & e_2 & 1        &   & \vdots \\
0 & 1 &  & \ddots  &   0 \\
\vdots  & & \ddots & ~\ddots~  &  ~1~  \\
0 &  \cdots  & 0 & 1 & e_n
\end{pmatrix}.
\end{equation}
where $\{e_{\ell}\}$ are determined by the ratio $p/q$ in the continuous fraction as 
\begin{equation*}
    \frac{p}{q} = [e_1, \ldots, e_{n}] =  e_1-\cfrac{1}{e_2-\cfrac{1}{\ddots e_{n-1}-\cfrac{1}{e_n}}}\;.
\end{equation*}
The fan of $X_{p,q}$ can be generated by the set of vectors $v_{\ell} \in N$ with $\ell=0,1,\ldots,n$. Here $v_0=(0,1)$ and $v_L=(p,-q)$. The others can be calculated recursively from the relation $v_{\ell+1}+v_{\ell-1}=e_\ell v_\ell$.

Consider a torus action on $X_{p,q}$ with $(z_1,z_2)\to (e^{i\epsilon_1} z_1, e^{i\epsilon_2} z_2)$. 
In terms of the invariant variables $w_1=z_1^p$ and $w_2=z_2/z_1^q$, the weights are shifted to
\begin{equation*}
    \epsilon \to M \epsilon \;, \qquad M=\begin{pmatrix}
p & 0\\
-q&1 \end{pmatrix}~.
\end{equation*}
By the equation (\ref{eqn:v2alpha}), the corresponding weights on each patch are 
\begin{equation} \label{eqn:v2alphaJung}
    \epsilon_1^{\ell} = -\det(v_{\ell+1}, M \epsilon), \qquad \epsilon_2^{\ell} = \det(v_{\ell}, M \epsilon)\;.
\end{equation}

\paragraph{Example: $\cO_{\bP^1}(-p)$}

This is the non-compact 4-manifold $X_{p,1}$ obtained from the resolution of toric singularities $\bC^2/\bZ_p$ with the action
\begin{equation*}
    ( z_1,z_2 ) \to e^{\frac{2\pi i}{p}} ( z_1, z_2 ) \;. 
\end{equation*}
The set of vectors of $X_{p,1}$ are 
\begin{equation*}
    v_0 = (0,1), \quad
    v_1 = (1,0), \quad
    v_2 = (p,-1) \;,
\end{equation*}
which implies that there is one exceptional divisor at the origin with self-intersection $e_1 = p$.

Given a torus action on $\bC^2$ with weights $\epsilon_{1,2}$, by the equation (\ref{eqn:v2alphaJung}), the corresponding weights on the patches are 
\begin{equation*}
   \alpha_1 = \frac{p\alpha}{1-\alpha},  \qquad \alpha_2 = \frac{\alpha-1}{p},
\end{equation*}
where $\alpha=\epsilon_2/\epsilon_1$. 
Using the localization formula (\ref{eqn:geo_formula}), the equivariant Euler characteristic and signature are 
\begin{equation*}
    \tilde{\chi}(\cO_{\bP^1}(-p)) = 2, \quad \tilde{\sigma}(\cO_{\bP^1}(-p)) =\frac{1}{3p}\left( \alpha + \frac{1}{\alpha} -(p^2+2)\right).
\end{equation*}

\paragraph{Example: $A_{p-1}$ space}

This is the non-compact 4-manifold $X_{p,p-1}$ obtained from the resolution of toric singularities $\bC^2/\bZ_p$ with the action
\begin{equation*}
    ( z_1,z_2 ) \to  ( e^{2\pi i/p}z_1, e^{-2\pi i/p}z_2 ) \;.
\end{equation*}
The set of vectors of $X_{p,p-1}$ are $\{v_{\ell} = (\ell,1-\ell)\}$ $\ell=0,1,\ldots, p$, which implies that there are $(p-1)$ exceptional divisors after the resolution with self intersection $e_{\ell} = 2$.

Given a torus action on $\bC^2$ with weights $\epsilon_{1,2}$, by the equation (\ref{eqn:v2alphaJung}), the corresponding weights on the $p$ patches are 
\begin{equation*}
    \alpha_0 = \frac{\alpha-(p-1)}{p}, \quad
    \alpha_1 = \frac{2\alpha-(p-2)}{(p-1) - \alpha},  \quad \ldots\ldots,
    \alpha_{p-1} = \frac{p\alpha}{1-(p-1)\alpha},
\end{equation*}
where $\alpha=\epsilon_2/\epsilon_1$. 
The origin of each patch contributes one fixed point under the torus action. 
Using the localization formula (\ref{eqn:geo_formula}), the equivariant Euler characteristic and signature are 
\begin{equation*} 
    \tilde \chi (A_{p-1}) = p, \qquad \tilde \sigma(A_{p-1}) = \frac{1}{3p}(\alpha+\frac{1}{\alpha}+2-2p^2).
\end{equation*}

\subsection{Plumbing 4-manifolds}
\label{app:B2}

A large class of non-compact 4-manifolds can be constructed by gluing $n$ disk bundles, $D^2_i \to S^2_i$, with Euler characteristic $a_i \in \bZ$ over the two-sphere. 
By switching the role of the base and the fiber, one can build a simply connected 4-manifold \cite{Gadde:2013sca}. 
This process can be conveniently described with a plumbing graph $\Upsilon$ in a way that each vertex represents a disk bundle, the Euler number of the bundle assigns to the weight of the vertices, and an edge between two vertices indicates that the corresponding bundles are glued together. 
In particular, for 4-manifolds without 1-cycles, we will avoid plumbing graphs that have loops. 
Therefore, in what follows we typically assume that $\Upsilon$ is a tree.

\begin{figure}[ht]
    \centering
    \resizebox{.25\textwidth}{!}{%
    \begin{tikzpicture}
        \tikzstyle{vertices} = [circle,inner sep=1.2pt, draw, fill=black];
        \tikzstyle{edge} = [-, -latex];
        \node (a1) [vertices,   label=above:{ $a_1$}] {};
        \node (a2) [vertices,   right of=a1, label=above:{$a_2$}] {};
        \node (a3) [vertices,   right of=a2, label=above:{$a_{3}$}] {};
        \node (a4) [   right of=a3,] {$\ldots$};
        \node (a5) [vertices,  right of=a4,  label=above:{$a_{n}$}] {};
        \draw (a1) -- (a2) (a2) -- (a3) (a3) -- (a4) (a4) -- (a5);
    \end{tikzpicture}
    }
    \caption{The plumbing graph of the $A_n$ manifold.}
    \label{fig:AN_manifold}
\end{figure}
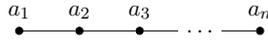

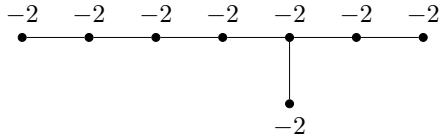
\begin{figure}[ht]
    \centering
    \resizebox{.4\textwidth}{!}{%
    \begin{tikzpicture}
        \tikzstyle{vertices} = [circle,inner sep=1.2pt, draw, fill=black];
        \node (a1) [vertices,   label=above:{$-2$}] {};
        \node (a2) [vertices,   right of=a1, label=above:{$-2$}] {};
        \node (a3) [vertices,   right of=a2, label=above:{$-2$}] {};
        \node (a4) [vertices,   right of=a3, label=above:{$-2$}] {};
        \node (a5) [vertices,   right of=a4, label=above:{$-2$}] {};
        \node (a6) [vertices,   right of=a5, label=above:{$-2$}] {};
        \node (a7) [vertices,   right of=a6, label=above:{$-2$}] {};
        \node (a8) [vertices,   below of=a5, label=below:{$-2$}] {};
        \draw (a1) --(a2) (a2)--(a3) (a3)--(a4) (a4)--(a5)  (a5)--(a6) (a6)--(a7) (a5)--(a8);
    \end{tikzpicture}
    }
    \caption{The plumbing graph of the $E_8$ manifold.}
    \label{fig:E8_manifold}
\end{figure}

Given a plumbing tree $\Upsilon$, the intersection form of the 4-manifold can be easily read from it by 
\begin{equation}
Q_{ij} \; = \;
\begin{cases}
a_i, & \text{if } i=j \\
1, & \text{if } i \text{ is connected to } j \text{ by an edge} \\
0, & \text{otherwise}
\end{cases}.
\label{Qplumbing}
\end{equation}
For example, the plumbing tree in Figure \ref{fig:AN_manifold} corresponds to
\begin{equation}
Q \; = \;
\begin{pmatrix}
a_1 & 1   & 0 & \cdots  & 0 \\
1   & a_2 & 1        &   & \vdots \\
0 & 1 &  & \ddots  &   0 \\
\vdots  & & \ddots & ~\ddots~  &  ~1~  \\
0 &  \cdots  & 0 & 1 & a_n
\end{pmatrix}.
\end{equation}
A further specialization to $(a_1, a_2, \ldots, a_n) = (-2, -2, \ldots, -2)$
for obvious reasons is usually referred to as $A_{n}$, whereas that in Figure \ref{fig:E8_manifold} is called $E_8$.

The plumbing graph are not unique. There are certain moves which relate
different presentations of the same 4-manifold. One of the important moves is the 
2-handle slide defined by the operation of sliding a
2-handle $i$ over a 2-handle $j$ \cite{Gadde:2013sca} 
\begin{equation} \label{eqn:2handleSilde}
    a_j  \mapsto  a_i + a_j \pm 2Q_{ij}, \qquad 
a_i  \mapsto  a_i
\end{equation}
where the $\pm$ sign is fixed by the choice of orientation (``$+$'' for handle addition and ``$-$'' for handle subtraction) and $Q_{ij}$ are the intersection number between different handles.

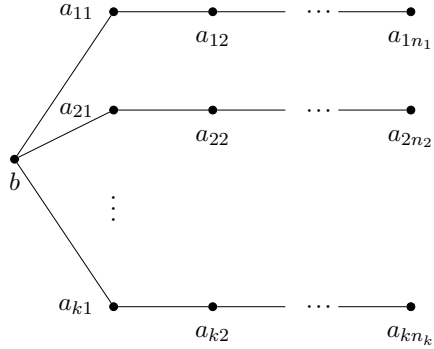
\begin{figure}
    \centering
    \resizebox{.4\textwidth}{!}{
    \begin{tikzpicture}[auto,rotate=90,transform shape]
        \tikzstyle{vertices} = [circle,inner sep=1.2pt, draw, fill=black];
        \node [vertices, label={[rotate=-90] left:$b$}] {} 
            child  {node [vertices, label={[label distance=1mm,rotate=-90] above:$\quad a_{k1}$ }] {}  
                child {node [vertices, label={[label distance=1mm,rotate=-90] left:$a_{k2}$}] {} 
                    child {node  {$\vdots$}  
                        child {node [vertices, label={[label distance=1mm,rotate=-90] left:$a_{kn_{k}}$}] {}}
                        }}}
            child {node {$\ldots$} edge from parent [white] }
            child  {node [vertices, label={[label distance=1mm,rotate=-90] above:$a_{21}$ }] {}  
                child {node [vertices, label={[label distance=1mm,rotate=-90] left:$a_{22}$}] {} 
                    child {node  {$\vdots$}  
                        child {node [vertices, label={[label distance=1mm,rotate=-90] left:$a_{2n_{2}}$}] {}}
                        }}}
            child  {node [vertices, label={[label distance=1mm,rotate=-90] above:$a_{11}$ }] {}  
                child {node [vertices, label={[label distance=1mm,rotate=-90] left:$a_{12}$}] {} 
                    child {node  {$\vdots$}  
                        child {node [vertices, label={[label distance=1mm,rotate=-90] left:$a_{1n_{1}}$}] {}}
                        }}};
    \end{tikzpicture}
    }
    \caption{A general plumbing tree.}
    \label{fig:general_plumbing_tree}
\end{figure}

A plumbing graph $\Upsilon$ of a non-compact 4-manifold $M_4$ also defines the boundary $\partial M_4=M_3$. 
For the most general plumbing tree $\Upsilon$ defined in Figure \ref{fig:general_plumbing_tree}, the corresponding boundary 3-manifold is a Seifert fibered homology 3-sphere $M_3 (b; (p_1,q_1),$ $\ldots, (p_k,q_k))$ with singular fibers of orders $p_i \ge 1$ where $-\frac{p_i}{q_i} = [a_{i1}, \ldots, a_{in_i}]$ are given by the following continued fractions 
\begin{equation}\label{eqn:Q_bnd}
    - \frac{p_i}{q_i} \; = \; a_{i1} - \cfrac{1}{a_{i2} - \cfrac{1}{\ddots - \cfrac{1}{a_{in_i}}}}.
\end{equation}
For example, the plumbing on $A_n$ has the Lens space boundary $M_3 = L(n+1,n)$, while the plumbing on $E_8$ has the Poincar\'e sphere boundary $M_3 = \Sigma (2,3,5)$. 
Notice that the representation of the boundary $M_3$ using $\Upsilon$ is not unique. There exists some moves on plumbing diagram called Kirby moves that do not change the boundary of the 4-manifolds. More detailed discussion on these moves can be found in \cite{Gadde:2013sca,Feigin:2018bkf}.

\bibliographystyle{unsrt}
\bibliography{references}

\begin{thebibliography}{10}

\bibitem{Gaiotto:2009we}
Davide Gaiotto.
\newblock {N=2 dualities}.
\newblock {\em JHEP}, 08:034, 2012.

\bibitem{Gaiotto:2009hg}
Davide Gaiotto, Gregory~W. Moore, and Andrew Neitzke.
\newblock {Wall-crossing, Hitchin systems, and the WKB approximation}.
\newblock {\em Adv. Math.}, 234:239--403, 2013.

\bibitem{Argyres:2007cn}
Philip~C. Argyres and Nathan Seiberg.
\newblock {S-duality in N=2 supersymmetric gauge theories}.
\newblock {\em JHEP}, 12:088, 2007.

\bibitem{Argyres:2007tq}
Philip~C. Argyres and John~R. Wittig.
\newblock {Infinite coupling duals of N=2 gauge theories and new rank 1
  superconformal field theories}.
\newblock {\em JHEP}, 01:074, 2008.

\bibitem{Alday:2009aq}
Luis~F. Alday, Davide Gaiotto, and Yuji Tachikawa.
\newblock {Liouville Correlation Functions from Four-dimensional Gauge
  Theories}.
\newblock {\em Lett. Math. Phys.}, 91:167--197, 2010.

\bibitem{Benini:2009mz}
Francesco Benini, Yuji Tachikawa, and Brian Wecht.
\newblock {Sicilian gauge theories and N=1 dualities}.
\newblock {\em JHEP}, 01:088, 2010.

\bibitem{Dimofte:2010tz}
Tudor Dimofte, Sergei Gukov, and Lotte Hollands.
\newblock {Vortex Counting and Lagrangian 3-manifolds}.
\newblock {\em Lett. Math. Phys.}, 98:225--287, 2011.

\bibitem{Dimofte:2011jd}
Tudor Dimofte and Sergei Gukov.
\newblock {Chern-Simons Theory and S-duality}.
\newblock {\em JHEP}, 05:109, 2013.

\bibitem{Dimofte:2011ju}
Tudor Dimofte, Davide Gaiotto, and Sergei Gukov.
\newblock {Gauge Theories Labelled by Three-Manifolds}.
\newblock {\em Commun. Math. Phys.}, 325:367--419, 2014.

\bibitem{Gadde:2013sca}
Abhijit Gadde, Sergei Gukov, and Pavel Putrov.
\newblock {Fivebranes and 4-manifolds}.
\newblock {\em Prog. Math.}, 319:155--245, 2016.

\bibitem{Gukov:2016gkn}
Sergei Gukov, Pavel Putrov, and Cumrun Vafa.
\newblock {Fivebranes and 3-manifold homology}.
\newblock {\em JHEP}, 07:071, 2017.

\bibitem{Zafrir:2015rga}
Gabi Zafrir.
\newblock {Brane webs, $5d$ gauge theories and $6d$ $\mathcal{N}=(1,0)$
  SCFT's}.
\newblock {\em JHEP}, 12:157, 2015.

\bibitem{Jefferson:2018irk}
Patrick Jefferson, Sheldon Katz, Hee-Cheol Kim, and Cumrun Vafa.
\newblock {On Geometric Classification of 5d SCFTs}.
\newblock {\em JHEP}, 04:103, 2018.

\bibitem{Bhardwaj:2019fzv}
Lakshya Bhardwaj, Patrick Jefferson, Hee-Cheol Kim, Houri-Christina Tarazi, and
  Cumrun Vafa.
\newblock {Twisted Circle Compactifications of 6d SCFTs}.
\newblock 9 2019.

\bibitem{Braun:2021lzt}
Andreas~P. Braun, Jin Chen, Babak Haghighat, Marcus Sperling, and Shuhang Yang.
\newblock {Fibre-base duality of 5d KK theories}.
\newblock {\em JHEP}, 05:200, 2021.

\bibitem{Gaiotto:2015una}
Davide Gaiotto and Hee-Cheol Kim.
\newblock {Duality walls and defects in 5d $ \mathcal{N}=1 $ theories}.
\newblock {\em JHEP}, 01:019, 2017.

\bibitem{Razamat:2016dpl}
Shlomo~S. Razamat, Cumrun Vafa, and Gabi Zafrir.
\newblock {4d $ \mathcal{N}=1 $ from 6d (1, 0)}.
\newblock {\em JHEP}, 04:064, 2017.

\bibitem{Bah:2017gph}
Ibrahima Bah, Amihay Hanany, Kazunobu Maruyoshi, Shlomo~S. Razamat, Yuji
  Tachikawa, and Gabi Zafrir.
\newblock {4d $ \mathcal{N}=1 $ from 6d $ \mathcal{N}=\left(1,0\right) $ on a
  torus with fluxes}.
\newblock {\em JHEP}, 06:022, 2017.

\bibitem{Kim:2017toz}
Hee-Cheol Kim, Shlomo~S. Razamat, Cumrun Vafa, and Gabi Zafrir.
\newblock {E-String Theory on Riemann Surfaces}.
\newblock {\em Fortsch. Phys.}, 66(1):1700074, 2018.

\bibitem{Kim:2018bpg}
Hee-Cheol Kim, Shlomo~S. Razamat, Cumrun Vafa, and Gabi Zafrir.
\newblock {D-type Conformal Matter and SU/USp Quivers}.
\newblock {\em JHEP}, 06:058, 2018.

\bibitem{Kim:2018lfo}
Hee-Cheol Kim, Shlomo~S. Razamat, Cumrun Vafa, and Gabi Zafrir.
\newblock {Compactifications of ADE conformal matter on a torus}.
\newblock {\em JHEP}, 09:110, 2018.

\bibitem{Razamat:2018gro}
Shlomo~S. Razamat and Gabi Zafrir.
\newblock {Compactification of 6d minimal SCFTs on Riemann surfaces}.
\newblock {\em Phys. Rev. D}, 98(6):066006, 2018.

\bibitem{Chen:2019njf}
Jin Chen, Babak Haghighat, Shuwei Liu, and Marcus Sperling.
\newblock {4d $N$=1 from 6d D-type $N$=(1,0)}.
\newblock {\em JHEP}, 01:152, 2020.

\bibitem{Razamat:2019mdt}
Shlomo~S. Razamat, Evyatar Sabag, and Gabi Zafrir.
\newblock {From 6d flows to 4d flows}.
\newblock {\em JHEP}, 12:108, 2019.

\bibitem{Cecotti:2011iy}
Sergio Cecotti, Clay Cordova, and Cumrun Vafa.
\newblock {Braids, Walls, and Mirrors}.
\newblock 10 2011.

\bibitem{Cordova:2013cea}
Clay Cordova and Daniel~L. Jafferis.
\newblock {Complex Chern-Simons from M5-branes on the Squashed Three-Sphere}.
\newblock {\em JHEP}, 11:119, 2017.

\bibitem{Yagi:2013fda}
Junya Yagi.
\newblock {3d TQFT from 6d SCFT}.
\newblock {\em JHEP}, 08:017, 2013.

\bibitem{Lee:2013ida}
Sungjay Lee and Masahito Yamazaki.
\newblock {3d Chern-Simons Theory from M5-branes}.
\newblock {\em JHEP}, 12:035, 2013.

\bibitem{Gukov:2015sna}
Sergei Gukov and Du~Pei.
\newblock {Equivariant Verlinde formula from fivebranes and vortices}.
\newblock {\em Commun. Math. Phys.}, 355(1):1--50, 2017.

\bibitem{Cordova:2016cmu}
Clay Cordova and Daniel~L. Jafferis.
\newblock {Toda Theory From Six Dimensions}.
\newblock {\em JHEP}, 12:106, 2017.

\bibitem{Gukov:2017kmk}
Sergei Gukov, Du~Pei, Pavel Putrov, and Cumrun Vafa.
\newblock {BPS spectra and 3-manifold invariants}.
\newblock {\em J. Knot Theor. Ramifications}, 29(02):2040003, 2020.

\bibitem{Cho:2020ljj}
Gil~Young Cho, Dongmin Gang, and Hee-Cheol Kim.
\newblock {M-theoretic Genesis of Topological Phases}.
\newblock {\em JHEP}, 11:115, 2020.

\bibitem{Apruzzi:2016nfr}
Fabio Apruzzi, Falk Hassler, Jonathan~J. Heckman, and Ilarion~V. Melnikov.
\newblock {From 6D SCFTs to Dynamic GLSMs}.
\newblock {\em Phys. Rev. D}, 96(6):066015, 2017.

\bibitem{Gukov:2018iiq}
Sergei Gukov, Du~Pei, Pavel Putrov, and Cumrun Vafa.
\newblock {4-manifolds and topological modular forms}.
\newblock {\em JHEP}, 05:084, 2021.

\bibitem{Pasquetti:2019hxf}
Sara Pasquetti, Shlomo~S. Razamat, Matteo Sacchi, and Gabi Zafrir.
\newblock {Rank $Q$ E-string on a torus with flux}.
\newblock {\em SciPost Phys.}, 8(1):014, 2020.

\bibitem{Assel:2022row}
Benjamin Assel, Yuji Tachikawa, and Alessandro Tomasiello.
\newblock {On ${\mathcal N}=4$ supersymmetry enhancements in three dimensions}.
\newblock 9 2022.

\bibitem{Maldacena_1997}
Juan Maldacena, Andrew Strominger, and Edward Witten.
\newblock Black hole entropy in m-theory.
\newblock {\em Journal of High Energy Physics}, 1997(12):002--002, dec 1997.

\bibitem{Gaiotto:2006wm}
Davide Gaiotto, Andrew Strominger, and Xi~Yin.
\newblock {The M5-Brane Elliptic Genus: Modularity and BPS States}.
\newblock {\em JHEP}, 08:070, 2007.

\bibitem{Dedushenko:2017tdw}
Mykola Dedushenko, Sergei Gukov, and Pavel Putrov.
\newblock {Vertex algebras and 4-manifold invariants}.
\newblock In {\em {Nigel Hitchin's 70th Birthday Conference}}, volume~1, pages
  249--318, 5 2017.

\bibitem{Feigin:2018bkf}
Boris Feigin and Sergei Gukov.
\newblock {VOA[$M_4$]}.
\newblock {\em J. Math. Phys.}, 61(1):012302, 2020.

\bibitem{Witten:1995ex}
Edward Witten.
\newblock {String theory dynamics in various dimensions}.
\newblock {\em Nucl. Phys. B}, 443:85--126, 1995.

\bibitem{Witten:1995zh}
Edward Witten.
\newblock {Some comments on string dynamics}.
\newblock In {\em {STRINGS 95: Future Perspectives in String Theory}}, pages
  501--523, 7 1995.

\bibitem{Witten:1995gx}
Edward Witten.
\newblock {Small instantons in string theory}.
\newblock {\em Nucl. Phys. B}, 460:541--559, 1996.

\bibitem{Seiberg:1996vs}
N.~Seiberg and Edward Witten.
\newblock {Comments on string dynamics in six-dimensions}.
\newblock {\em Nucl. Phys. B}, 471:121--134, 1996.

\bibitem{Morrison:1996pp}
David~R. Morrison and Cumrun Vafa.
\newblock {Compactifications of F theory on Calabi-Yau threefolds. 2.}
\newblock {\em Nucl. Phys. B}, 476:437--469, 1996.

\bibitem{Seiberg:1996qx}
Nathan Seiberg.
\newblock {Nontrivial fixed points of the renormalization group in
  six-dimensions}.
\newblock {\em Phys. Lett. B}, 390:169--171, 1997.

\bibitem{Hanany:1997gh}
Amihay Hanany and Alberto Zaffaroni.
\newblock {Branes and six-dimensional supersymmetric theories}.
\newblock {\em Nucl. Phys. B}, 529:180--206, 1998.

\bibitem{Morrison:2012np}
David~R. Morrison and Washington Taylor.
\newblock {Classifying bases for 6D F-theory models}.
\newblock {\em Central Eur. J. Phys.}, 10:1072--1088, 2012.

\bibitem{Heckman:2013pva}
Jonathan~J. Heckman, David~R. Morrison, and Cumrun Vafa.
\newblock {On the Classification of 6D SCFTs and Generalized ADE Orbifolds}.
\newblock {\em JHEP}, 05:028, 2014.
\newblock [Erratum: JHEP 06, 017 (2015)].

\bibitem{DelZotto:2014hpa}
Michele Del~Zotto, Jonathan~J. Heckman, Alessandro Tomasiello, and Cumrun Vafa.
\newblock {6d Conformal Matter}.
\newblock {\em JHEP}, 02:054, 2015.

\bibitem{Heckman:2014qba}
Jonathan~J. Heckman.
\newblock {More on the Matter of 6D SCFTs}.
\newblock {\em Phys. Lett. B}, 747:73--75, 2015.
\newblock [Erratum: Phys.Lett.B 808, 135675 (2020)].

\bibitem{Heckman:2015bfa}
Jonathan~J. Heckman, David~R. Morrison, Tom Rudelius, and Cumrun Vafa.
\newblock {Atomic Classification of 6D SCFTs}.
\newblock {\em Fortsch. Phys.}, 63:468--530, 2015.

\bibitem{Gukov:1999ya}
Sergei Gukov, Cumrun Vafa, and Edward Witten.
\newblock {CFT's from Calabi-Yau four folds}.
\newblock {\em Nucl. Phys. B}, 584:69--108, 2000.
\newblock [Erratum: Nucl.Phys.B 608, 477--478 (2001)].

\bibitem{Bonelli:2012ny}
Giulio Bonelli, Kazunobu Maruyoshi, Alessandro Tanzini, and Futoshi Yagi.
\newblock {N=2 gauge theories on toric singularities, blow-up formulae and
  W-algebrae}.
\newblock {\em JHEP}, 01:014, 2013.

\bibitem{Alim:2010cf}
Murad Alim, Babak Haghighat, Michael Hecht, Albrecht Klemm, Marco Rauch, and
  Thomas Wotschke.
\newblock {Wall-crossing holomorphic anomaly and mock modularity of multiple
  M5-branes}.
\newblock {\em Commun. Math. Phys.}, 339(3):773--814, 2015.

\bibitem{Haghighat:2011xx}
Babak Haghighat and Stefan Vandoren.
\newblock {Five-dimensional gauge theory and compactification on a torus}.
\newblock {\em JHEP}, 09:060, 2011.

\bibitem{Haghighat:2012bm}
Babak Haghighat, Jan Manschot, and Stefan Vandoren.
\newblock {A 5d/2d/4d correspondence}.
\newblock {\em JHEP}, 03:157, 2013.

\bibitem{Horava:1995qa}
Petr Horava and Edward Witten.
\newblock {Heterotic and type I string dynamics from eleven-dimensions}.
\newblock {\em Nucl. Phys. B}, 460:506--524, 1996.

\bibitem{Morrison:2014lca}
David~R. Morrison and Washington Taylor.
\newblock {Non-Higgsable clusters for 4D F-theory models}.
\newblock {\em JHEP}, 05:080, 2015.

\bibitem{Witten:1996md}
Edward Witten.
\newblock {On flux quantization in M theory and the effective action}.
\newblock {\em J. Geom. Phys.}, 22:1--13, 1997.

\bibitem{Delmastro:2019vnj}
Diego Delmastro and Jaume Gomis.
\newblock {Symmetries of Abelian Chern-Simons Theories and Arithmetic}.
\newblock {\em JHEP}, 03:006, 2021.

\bibitem{Ohmori:2014kda}
Kantaro Ohmori, Hiroyuki Shimizu, Yuji Tachikawa, and Kazuya Yonekura.
\newblock {Anomaly polynomial of general 6d SCFTs}.
\newblock {\em PTEP}, 2014(10):103B07, 2014.

\bibitem{Ohmori:2014pca}
Kantaro Ohmori, Hiroyuki Shimizu, and Yuji Tachikawa.
\newblock {Anomaly polynomial of E-string theories}.
\newblock {\em JHEP}, 08:002, 2014.

\bibitem{Heckman:2018jxk}
Jonathan~J. Heckman and Tom Rudelius.
\newblock {Top Down Approach to 6D SCFTs}.
\newblock {\em J. Phys. A}, 52(9):093001, 2019.

\bibitem{Harvey:1998bx}
Jeffrey~A. Harvey, Ruben Minasian, and Gregory~W. Moore.
\newblock {NonAbelian tensor multiplet anomalies}.
\newblock {\em JHEP}, 09:004, 1998.

\bibitem{Alday:2009qq}
Luis~F. Alday, Francesco Benini, and Yuji Tachikawa.
\newblock {Liouville/Toda central charges from M5-branes}.
\newblock {\em Phys. Rev. Lett.}, 105:141601, 2010.

\bibitem{Agarwal:2012bn}
Abhishek Agarwal and V.~P. Nair.
\newblock {Supersymmetry and Mass Gap in 2+1 Dimensions: A Gauge Invariant
  Hamiltonian Analysis}.
\newblock {\em Phys. Rev. D}, 85:085011, 2012.

\bibitem{Fucito:2006kn}
Francesco Fucito, Jose~F. Morales, and Rubik Poghossian.
\newblock {Instanton on toric singularities and black hole countings}.
\newblock {\em JHEP}, 12:073, 2006.

\bibitem{Griguolo:2006kp}
Luca Griguolo, Domenico Seminara, Richard~J. Szabo, and Alessandro Tanzini.
\newblock {Black holes, instanton counting on toric singularities and
  q-deformed two-dimensional Yang-Mills theory}.
\newblock {\em Nucl. Phys. B}, 772:1--24, 2007.

\bibitem{Dijkgraaf:2007fe}
Robbert Dijkgraaf and Piotr Sulkowski.
\newblock {Instantons on ALE spaces and orbifold partitions}.
\newblock {\em JHEP}, 03:013, 2008.

\bibitem{Cirafici:2009ga}
Michele Cirafici, Amir-Kian Kashani-Poor, and Richard~J. Szabo.
\newblock {Crystal melting on toric surfaces}.
\newblock {\em J. Geom. Phys.}, 61:2199--2218, 2011.

\bibitem{Gasparim:2009sns}
Elizabeth Gasparim and Chiu-Chu~Melissa Liu.
\newblock {The Nekrasov Conjecture for Toric Surfaces}.
\newblock {\em Commun. Math. Phys.}, 293:661--700, 2010.

\bibitem{Bonelli:2011jx}
Giulio Bonelli, Kazunobu Maruyoshi, and Alessandro Tanzini.
\newblock {Instantons on ALE spaces and Super Liouville Conformal Field
  Theories}.
\newblock {\em JHEP}, 08:056, 2011.

\bibitem{Bonelli:2011kv}
Giulio Bonelli, Kazunobu Maruyoshi, and Alessandro Tanzini.
\newblock {Gauge Theories on ALE Space and Super Liouville Correlation
  Functions}.
\newblock {\em Lett. Math. Phys.}, 101:103--124, 2012.

\bibitem{Ito:2011mw}
Yuto Ito.
\newblock {Ramond sector of super Liouville theory from instantons on an ALE
  space}.
\newblock {\em Nucl. Phys. B}, 861:387--402, 2012.

\bibitem{Alfimov:2011ju}
M.~N. Alfimov and G.~M. Tarnopolsky.
\newblock {Parafermionic Liouville field theory and instantons on ALE spaces}.
\newblock {\em JHEP}, 02:036, 2012.

\bibitem{Bruzzo:2009uc}
Ugo Bruzzo, Rubik Poghossian, and Alessandro Tanzini.
\newblock {Poincare polynomial of moduli spaces of framed sheaves on (stacky)
  Hirzebruch surfaces}.
\newblock {\em Commun. Math. Phys.}, 304:395--409, 2011.

\bibitem{Bruzzo:2013daa}
Ugo Bruzzo, Mattia Pedrini, Francesco Sala, and Richard~J. Szabo.
\newblock {Framed sheaves on root stacks and supersymmetric gauge theories on
  ALE spaces}.
\newblock {\em Adv. Math.}, 288:1175--1308, 2016.

\bibitem{Bruzzo:2014jza}
Ugo Bruzzo, Francesco Sala, and Richard~J. Szabo.
\newblock {${\mathcal{N} = 2}$ Quiver Gauge Theories on A-type ALE Spaces}.
\newblock {\em Lett. Math. Phys.}, 105(3):401--445, 2015.

\bibitem{Bawane:2014uka}
Aditya Bawane, Giulio Bonelli, Massimiliano Ronzani, and Alessandro Tanzini.
\newblock {$\mathcal{N}=2$ supersymmetric gauge theories on $S^2\times S^2$ and
  Liouville Gravity}.
\newblock {\em JHEP}, 07:054, 2015.

\bibitem{Bershtein:2015xfa}
Mikhail Bershtein, Giulio Bonelli, Massimiliano Ronzani, and Alessandro
  Tanzini.
\newblock {Exact results for $ \mathcal{N} $ = 2 supersymmetric gauge theories
  on compact toric manifolds and equivariant Donaldson invariants}.
\newblock {\em JHEP}, 07:023, 2016.

\bibitem{Bershtein:2016mxz}
Mikhail Bershtein, Giulio Bonelli, Massimiliano Ronzani, and Alessandro
  Tanzini.
\newblock {Gauge theories on compact toric surfaces, conformal field theories
  and equivariant Donaldson invariants}.
\newblock {\em J. Geom. Phys.}, 118:40--50, 2017.

\bibitem{Hosseini:2020vgl}
Seyed~Morteza Hosseini, Kiril Hristov, Yuji Tachikawa, and Alberto Zaffaroni.
\newblock {Anomalies, Black strings and the charged Cardy formula}.
\newblock {\em JHEP}, 09:167, 2020.

\bibitem{Vafa:2015euh}
Cumrun Vafa.
\newblock {Fractional Quantum Hall Effect and M-Theory}.
\newblock 11 2015.

\bibitem{Alfimov:2013cqa}
M.~N. Alfimov, A.~A. Belavin, and G.~M. Tarnopolsky.
\newblock {Coset conformal field theory and instanton counting on
  $C^{2}/Z_{p}$}.
\newblock {\em JHEP}, 08:134, 2013.

\bibitem{Chang:2011vka}
Chi-Ming Chang and Xi~Yin.
\newblock {Correlators in $W_N$ Minimal Model Revisited}.
\newblock {\em JHEP}, 10:050, 2012.

\bibitem{Mitev:2017jqj}
Vladimir Mitev and Elli Pomoni.
\newblock {2D CFT blocks for the 4D class $\mathcal{S}_k$ theories}.
\newblock {\em JHEP}, 08:009, 2017.

\bibitem{Nishioka:2011jk}
Tatsuma Nishioka and Yuji Tachikawa.
\newblock {Central charges of para-Liouville and Toda theories from
  M-5-branes}.
\newblock {\em Phys. Rev. D}, 84:046009, 2011.

\bibitem{Cordova:2013bea}
Clay Cordova and Daniel~L. Jafferis.
\newblock {Five-Dimensional Maximally Supersymmetric Yang-Mills in Supergravity
  Backgrounds}.
\newblock {\em JHEP}, 10:003, 2017.

\bibitem{Cordes:1994fc}
Stefan Cordes, Gregory~W. Moore, and Sanjaye Ramgoolam.
\newblock {Lectures on 2-d Yang-Mills theory, equivariant cohomology and
  topological field theories}.
\newblock {\em Nucl. Phys. B Proc. Suppl.}, 41:184--244, 1995.

\bibitem{Pestun:2016zxk}
Vasily Pestun et~al.
\newblock {Localization techniques in quantum field theories}.
\newblock {\em J. Phys. A}, 50(44):440301, 2017.

\bibitem{Bah:2019rgq}
Ibrahima Bah, Federico Bonetti, Ruben Minasian, and Emily Nardoni.
\newblock {Anomalies of QFTs from M-theory and Holography}.
\newblock {\em JHEP}, 01:125, 2020.

\end{thebibliography}

\end{document}